\documentclass[twocolumn,showpacs,nofootinbib,preprintnumbers,amsmath,amssymb]{revtex4-1}
\usepackage{epsfig}
\pdfoutput=1
\usepackage{amsmath,amssymb,amsthm,amsfonts}

\usepackage{mathrsfs}
\usepackage{slashed}
\usepackage{graphicx}
\usepackage{subfigure}
\usepackage{color,rotating}
\usepackage{dsfont}
\usepackage{setspace}
\usepackage{verbatim}
\usepackage{fancyhdr}
\usepackage{hyperref}
\usepackage{mathtools}
\usepackage{array}
\usepackage{tikz}
\usepackage[normalem]{ulem} %emphasize weiterhin kursiv
\usepackage{ulem} %\emph{Text}: Text wird unterstrichen
\usepackage{soul}
\usepackage{cancel}

\newcommand{\beq}{\begin{equation}}
\newcommand{\eeq}{\end{equation}}
\newcommand{\beqa}{\begin{eqnarray}}
\newcommand{\eeqa}{\end{eqnarray}}
\newcommand{\bfc}{\begin{figure}[h!]\begin{center}}
\newcommand{\efc}{\end{center}\end{figure}}

\newcommand{\bea}{\begin{eqnarray}}
\newcommand{\eea}{\end{eqnarray}}
\newcommand{\Eq}[1]{Eq.~(\ref{#1})}

\def\eq#1{(\ref{#1})}
\def\Eq#1{Eq.~(\ref{#1})}
\newcommand {\apgt} {\ {\raise-.5ex\hbox{$\buildrel>\over\sim$}}\ }
\newcommand {\aplt} {\ {\raise-.5ex\hbox{$\buildrel<\over\sim$}}\ }

\def\s0#1#2{\mbox{\small{$ \frac{#1}{#2} $}}}
\def\0#1#2{\frac{#1}{#2}}

\newcommand{\I}{\mathrm{i}}
\newcommand{\be}{\begin{eqnarray}}
\newcommand{\ee}{\end{eqnarray}}

\usepackage{color}

\def\eq#1{(\ref{eq:#1})}
\def\Eq#1{Eq.~(\ref{eq:#1})}
\def\Eqs#1{Eqs.~(\ref{eq:#1})}

\newcommand{\Fig}[1]{Fig.~\ref{fig:#1}}

\newcommand{\Sect}[1]{Sect.~\ref{sec:#1}}
\newcommand{\sect}[1]{\ref{sec:#1}}

\newcommand{\App}[1]{App.~\ref{app:#1}}
\newcommand{\app}[1]{\ref{app:#1}}

\newcommand{\weg}[1]{}

\renewcommand{\vec}[1]{\mathbf{#1}}

\newcommand{\nuk}[2]{\tilde{\nu}_{#1}^{#2}}
\newcommand{\nukk}{\nuk{k}{}}
\newcommand{\nukp}{\nuk{p}{}}
\newcommand{\nukps}{\nuk{p}{2}}
\newcommand{\nukq}{\nuk{q}{}}
\newcommand{\nukr}{\nuk{r}{}}
\newcommand{\T}{T(\hat{q}){}}
\newcommand{\twidlT}{\tilde{T}(\epsilon){}}

\begin{document}

%=======================================================================
\title{Anomalous scaling at non-thermal fixed points\\ of Burgers' and Gross-Pitaevskii turbulence
}

\author{Steven Mathey}
\email{s.mathey@thphys.uni-heidelberg.de}
\author{Thomas Gasenzer}
\email{t.gasenzer@thphys.uni-heidelberg.de}
\author{Jan M.~Pawlowski}
\email{j.pawlowski@thphys.uni-heidelberg.de}
\affiliation{Institut f\"ur Theoretische
  Physik, Universit\"at Heidelberg, Philosophenweg 16, 69120
  Heidelberg, Germany, \\
  Heidelberg Center for Quantum Dynamics, Universit\"at Heidelberg, INF 226, 69120 Heidelberg, Germany,\\
ExtreMe Matter Institute EMMI, GSI,
  Planckstra{\ss}e 1, D-64291 Darmstadt, Germany
  }
\date{\today}

%=======================================================================
    \begin{abstract}
Scaling in the dynamical properties of complex many-body systems has been of strong interest since turbulence phenomena became the subject of systematic mathematical studies.
In this article, dynamical critical phenomena far from equilibrium are investigated with  functional renormalisation group equations.
The focus is set on scaling solutions of the stochastic driven-dissipative Burgers equation and their relation to 
solutions known in the literature for Burgers and Kardar-Parisi-Zhang dynamics.
We furthermore relate superfluid as well as acoustic turbulence described by the Gross-Pitaevskii model to known analytic and numerical results for scaling solutions.
In this way, the canonical Kolmogorov exponent $5/3$ for the energy cascade in superfluid turbulence is obtained analytically.
We also get first results for anomalous exponents of acoustic and quantum turbulence.
These are consistent with existing experimental data.
Our results should be relevant for future experiments with, e.g., exciton-polariton condensates in solid-state systems as well as with ultra-cold atomic gases.
    \end{abstract}


\maketitle

%=======================================================================
%=======================================================================
\section{Introduction}
The concept of scaling has been of tremendous interest since the days of Galilei to whom the insight is attributed that the effects of physical laws may change considerably under a mere rescaling of object sizes \cite{Peterson2002a.AmJPh70.575}.
Scale transformations became important in studies of turbulent flow, beginning with Reynolds' work and culminating in Kolmogorov's seminal papers of 1941  \cite{Kolmogorov1941a,*Kolmogorov1941b,*Kolmogorov1941c,Frisch2004a}.
In the context of phase transitions between equilibrium states of matter, scaling behaviour is a hallmark of criticality and the appearance of macroscopic structure independent of microscopic details.
When a system near a phase transition is driven away from equilibrium, critical properties also show up in the ensuing dynamical evolution \cite{Hohenberg1977a}.
The correlation length grows large, and relaxation time scales diverge. 
Critical dynamics away from thermal equilibrium has been studied extensively in the physics of non-linear systems, in particular in the context of pattern formation \cite{cross1993a}, phase-ordering kinetics \cite{bray1994,*damle1996}, and turbulence \cite{nelkin1974turbulence,*Eyink1994a,*bramwell1998universality,*eckhardt2011critical}, but the full structure of non-thermal criticality is still far from being satisfactorily understood.

Here, we study, by means of functional renorma\-li\-sation-group methods, non-thermal critical states of driven and dissipative hydrodynamics in view of possible scaling.
We consider classical Burgers' turbulence  \cite{Burgers1939a,Frisch2000a,Bec2007a}, and relate our results to scaling solutions of the Kardar-Parisi-Zhang (KPZ) equation \cite{Kardar1986a} as well as to quantum turbulence described by the Gross-Pitaevskii (GP) model \cite{Gross1961a,*Pitaevskii1961a}.
We study, specifically, possible scalings at fixed points of the driven-dissipative Burgers equation and discuss their relevance in the context of turbulent cascades.
By comparing with semi-classical simulations of the GP equation, we show that critical  exponents known for the KPZ equation can be used to quantify anomalous scaling of acoustic turbulence in a superfluid.

The dynamics of many-body systems driven out of equilibrium can become stationary by means of dissipation and be characterised by a local flux of energy, both in position and momentum space.  
Such dynamics has many realisations in nature since it is virtually impossible to fully suppress contact to the environment in any realistic setting.  
If the driving, the flux of energy in momentum space, and the dissipation have suitable characteristics, the stationary dynamics can exhibit scale invariance and universality distinct from any thermal equilibrium state.  
Examples of such driven-dissipative stationary systems are realised in a wide range of systems, from exciton-polariton condensates \cite{Carusotto2013a,weisbuch1992a,Kasprzak2006a,lai2007a,*deng2007a,%
amo2009a,lagoudakis2008a,*Lagoudakis2009a,Amo2011a,*hivet2012a}, through pattern formation in non-linear media \cite{aranson2002a,cross1993a}  all the way to classical hydrodynamic turbulence \cite{Kolmogorov1941a,*Kolmogorov1941b,*Kolmogorov1941c,Frisch2004a}.

In this article, we study scaling solutions of the stochastic Burgers equation \cite{Burgers1939a,Frisch2000a,Bec2007a},
which is a model for fully compressible fluid dynamics. 
We compute stationary correlation functions of Burgers turbulence driven by a random Gaussian forcing.
We set up functional renormalisation group (RG) flow equations to look for non-perturbative fixed points, applying ideas from Refs.~\cite{Pawlowski:2003XX,Gasenzer:2008zz,Gasenzer:2010rq} in the context of classical hydrodynamic turbulence \cite{Collina1997a,Barbi2010a,Mejiamonasterio2012a}.
For this, we take momentum dependence of low-order correlations into account.

We identify a range of critical scalings corresponding to solutions regular in the ultraviolet.
These scalings comprise that at known fixed points of the KPZ equation \cite{Kardar1986a}.
The respective exponents are shown to corroborate numerical results for sound-wave turbulence in GP superfluids, giving rise to first estimates of anomalous exponents at non-thermal fixed points  \cite{Berges:2008wm,Berges:2008sr,Scheppach:2009wu}.
Fixed points outside this range, which require a UV regulator to be implemented in the integrals, can, to a certain extent, be related to direct cascades of energy.
Kolmogorov scaling of incompressible fluids as well as superfluid turbulence \cite{Vinen2006a,Maurer1998,Walmsley2007a,*Walmsley2008a,Nazarenko2001a,Araki2002a,Kobayashi2005a,Volovik2004a,Kozik2009a,Tsubota2008a,*Tsubota2010a,*Numasato2010a} belong to this regime.

For the GP model, non-thermal fixed points are known to exist \cite{Scheppach:2009wu} which are related to strong phase excitations, including ensembles of quasi-topological defects such as vortices \cite{Nowak:2010tm,Nowak:2011sk,Schole:2012kt,Nowak:2013juc} or solitons \cite{Schmidt:2012kw,Karl:2013kua}, as well as local density depressions and sound-wave turbulence \cite{Nowak:2011sk}. 
Here, we use an additional constraint set by Galilei invariance to analytically obtain the canonical scaling laws at these fixed points: 
While the quasi-particle cascade exhibits the known $p^{-1}$ scaling of random vortex-antivortex ensembles, the energy cascade exhibits Kolmogorov-$p^{-5/3}$ scaling known from simulations of the GPE \cite{Araki2002a,Kobayashi2005a}.
Comparing simulation results for superfluid turbulence \cite{Araki2002a,Kobayashi2005a} and non-thermal fixed points \cite{Nowak:2010tm,Nowak:2011sk,Schole:2012kt} with analytic predictions of the present work as well as of the strong-wave-turbulence analysis  \cite{Scheppach:2009wu} we conjecture that anomalous exponents for the respective scaling solutions of the GP model are close to zero.
Finally, we find signatures of an additional yet unknown fixed point which is associated with spatially non-local forcing.
For $d=1$ our results are in agreement with perturbative \cite{Medina1989a} and non-perturbative \cite{Canet2010a,Canet2011a,Kloss2012a,Kloss2013a} RG calculations.

Our results should be of relevance for future experiments with driven-dissipative systems such as exciton polaritons \cite{Carusotto2013a,weisbuch1992a,Kasprzak2006a,lai2007a,*deng2007a,%
amo2009a,lagoudakis2008a,*Lagoudakis2009a,Amo2011a,*hivet2012a} as well as ultracold atomic gases \cite{Anderson1995a,*Weiner1999a,*Kwon2014} which have the potential to measure particle number distributions in momentum space and thus observe power laws directly.

Our paper is organised as follows:
In \Sect{turbulence_in_DDS}, we discuss driven-dissipative dynamics of the stochastic Burgers and GP equations. 
In \Sect{RGapproach}, we introduce the functional RG approach before setting up flow equations in \Sect{approximation}. 
The RG fixed point equations are derived and discussed in \Sect{fixed_point_equations}.
Analytical constraints on their properties are presented in \Sect{constraints}. 
In Sect.~\sect{comparisons}, we discuss the physical implications of our results for classical and quantum turbulence. 

%=======================================================================
%=======================================================================
\section{Turbulence in driven dissipative systems}
\label{sec:turbulence_in_DDS}
%
%=======================================================================
\subsection{Burgers and KPZ turbulence}
\label{sec:burgers_turbulence}
In this article, we study scaling solutions of the stochastic Burgers equation \cite{Burgers1939a,Frisch2000a,Bec2007a},
\begin{align}
\partial_t {\bf v} + ({\bf v} \cdot \boldsymbol{\nabla}) {\bf v} - \nu \nabla^2 {\bf v} = \mathbf{f}.
\label{eq:Burgers}
\end{align}
${\bf  v}$ is the position and time dependent velocity field and $\nu$ is the kinematic viscosity.  
$\mathbf{f}$ is a force with zero average, $\langle \mathbf{f}\rangle = 0$, and Gaussian fluctuations {$\langle f_{i}(t,\mathbf{x}) f_{j}(t',\mathbf{x'})\rangle = \delta(t-t') F_{ij}(\left|\mathbf{x}-\mathbf{x'}\right|)$}. Latin indices denote spatial dimensions.
To distinguish different types of forcing we choose the power-law ansatz
\begin{align}
\langle f_i(\omega,\mathbf{p}) f_{j}(\omega',\mathbf{p'})\rangle = \delta_{ij} \, \delta(\omega+\omega') \, \delta(\mathbf{p}+\mathbf{p'}) \, F \, p^{\beta}
\label{eq:forcingcorr}
\end{align}
{for the force correlator in Fourier-space, where $p=|\mathbf{p}|$.}
Hence, the exponent $\beta$ determines the degree of non-locality of the forcing. 
For $\beta > 0$ the energy is mainly injected into the UV modes while for $\beta<0$ the forcing acts on large scales. The case $\beta = 0$ corresponds to a forcing delta correlated in space.

Burgers' equation is equivalent to the Navier-Stokes equation if the equation of state is assumed to impose a constant pressure, $P = \ $const.~\cite{guyon2001a}.
For applications of Burgers' equation see Ref.~\cite{Bec2007a} and references therein. 
The irrotationally forced Burgers equation can be mapped onto the Kardar-Parisi-Zhang (KPZ) equation \cite{Kardar1986a},
\begin{align}
 {\partial_t \theta + \frac{\lambda}{2} \left(\boldsymbol{\nabla} \theta \right)^2 = \nu \nabla^2 \theta,}
\end{align}
with $\mathbf{v} = \boldsymbol{\nabla} \theta$.
The KPZ equation is typically used to describe non-linear interface growth but can also be applied to the dynamics of phase fluctuations in an ultracold Bose gas described by the stochastic GP model \cite{Carusotto2013a,Altman2013a}, or to directed polymers in random media \cite{Huse1985a,Kardar1987a,Bouchaud1995a}.
In the limit $\nu \rightarrow 0$,  Burgers' equation  possesses shock-wave solutions with discontinuities in the velocity field \cite{Grafke2013a,Mesterhazy2013b}.
These shocks can also appear in the GP model but, due to the definition of the phase on a compact circle, lead to the creation of (quasi) topological defects, e.g., dissolve into soliton trains \cite{Dutton2001a,Kevrekidis2008a}.

An extensive range of studies of critical dynamics exists for the models studied here.
To characterize a turbulent state, an important quantity is the second moment of the velocity increment,
\begin{align}
 \Delta v (\tau,\mathbf{r}) 
 = \langle \left[ \mathbf{v}(t+\tau,\mathbf{x} + \mathbf{r}) - \mathbf{v}(t,\mathbf{x}) \right]^2 \rangle,
 \label{eq:Deltav}
\end{align}
which takes the scaling form
\begin{align}
\Delta v (\tau,\mathbf{r}) = r^{2(\chi-1)} g\left(\tau/r^z\right),
\label{eq:DeltavScaling}
\end{align}
with $r=|\mathbf{r}|$ and roughness and dynamical critical exponents $\chi$ and $z$, respectively.
We will in particular consider the kinetic energy spectrum
\begin{align}
\epsilon_{\text{kin}}(\mathbf{p})
= \frac{1}{2} \int_{\omega} \, \langle \mathbf{v}(\omega,\mathbf{p}) \cdot \mathbf{v}(-\omega,-\mathbf{p})\rangle ,
\label{eq:kin_energy_1}
\end{align}
and derive the scaling exponent $\xi$ defined by 
\begin{align}
 \epsilon_{\text{kin}}(s|\mathbf{p}|)
 &= s^{-\xi}\epsilon_{\text{kin}}(|\mathbf{p}|).
 \label{eq:xi}
\end{align}
Here and in the following we use the short-hand notation $\int_{t,\mathbf{x}}=\int\mathrm{d}t\,\mathrm{d}^{d}x$, $\int_{\omega,\mathbf{p}}=(2\pi)^{-d-1}\int\mathrm{d}\omega\,\mathrm{d}^{d}p$. 

The values of $\chi$ and $z$ are, so far, only known for spatial dimension $d=1$ and specific choices of the forcing and initial conditions, see Refs.~\cite{Sasamoto2010a,Corwin2012a}. 
For $d\geq2$, the analytical studies  \cite{sun1994a,Frey1994a,Gurarie1996a,Fedorenko2013a} assume that the forcing (and/or velocity field) is the gradient of a potential and exploit the mapping to the KPZ equation.
Numerical studies exist for the one-dimensional case \cite{Hayot1996a,Bec2007a,Grafke2013a,Mesterhazy2013b}.

Scaling of correlation functions as described by the KPZ equation has been studied in Ref.~\cite{Medina1989a} by means of a one-loop perturbative RG calculation. 
A scaling $\langle \left|f(\omega,\mathbf{p})\right|^2 \rangle \sim p^{\beta}$ of the force correlator is assumed, and the scaling exponent $\chi$ of the correlation function \eq{DeltavScaling} is uniquely related to the exponent $\beta$ of the forcing, for $0<\beta<2$ and $d=1$. 
Outside this range, non-perturbative effects become important and a more sophisticated approach is necessary.

An exact expression for the time-dependent velocity field probability distribution was obtained in the case $\beta = 2$ and $d=1$. 
For reviews, see \cite{Sasamoto2010a,Corwin2012a} and references therein. 
Its asymptotic limit confirms the predictions of Refs.~\cite{Forster1977a,Kardar1986a} concerning the scaling exponents and yields $\chi = 1/2$ and $z=3/2$.
However, due to the limitations of perturbative methods, possible scaling forms of the correlator are unknown in $d>1$ dimensions. 
The question of the existence of an upper critical dimension is still open.

Non-perturbative approximations provide insights into the behaviour of the KPZ equation for $d>1$. 
Most of the literature concentrate on the case $\beta = 2$, corresponding to white-noise forcing in space. 
In this context, predictions for the scaling exponents, scaling functions and upper critical dimension have been made, using, e.g., the mode coupling approximation \cite{vanBeijeren1985a,*Bouchaud1993a,*Frey1996a,*Colaiori2001a}, the self-consistent expansion \cite{schwartz1992a,*Schwartz2008a}, or the weak-noise scheme \cite{Fogedby2001a,*Fogedby2005a,*Fogedby2006a}.
The case $\beta < 0$ of a forcing that is concentrated on large scales was tackled in Ref.~\cite{Bouchaud1995a} by means of a replica-trick approach being exact in the limit $d \to \infty$, and bi-fractal scaling of the velocity increments was obtained.
The tails of the probability distribution of velocity differences were addressed in \cite{Polyakov1995a}
using an operator product expansion in $d=1$, and in \cite{Gurarie1996a,
Grafke2013a} within an instanton approach. 
Decaying Burgers turbulence was studied in \cite{Fedorenko2013a}. 

The stochastic KPZ equation, for the case $\beta = 2$, has been studied within the functional RG framework in Refs.~\cite{Canet2010a,Canet2011a,Kloss2012a}.
Non-perturbative RG fixed points were found for $d\leq3$, and scaling exponents and functions were computed.
Furthermore, the perturbative results of \cite{Kardar1986a} were recovered, and the form of the scaling function $g(x)$ derived compared well with the exact results of Refs.~\cite{Sasamoto2010a,Corwin2012a}.

%=======================================================================
\subsection{Driven dissipative Gross-Pitaevskii systems}
\label{sec:B2GPE}
Driven-dissipative superfluid dilute Bose gases can be described in terms of the stochastic Gross-Pitaevskii equation (SGPE),
\begin{align}
 i \partial_t \psi = \left[-\left(\frac{1}{2m}-i\nu\right) \nabla^2 - \mu + g \left|\psi \right|^2 \right] \psi+\zeta.
\label{eq:GPE}
\end{align}
Here and in the following,  $\hbar = 1$.
Driven-dissipative non-linear equations of the type \eq{GPE} have been studied in the literature, see, e.g., \cite{ Pismen1999a,bohr2005dynamical,aranson2002a,Kobayashi2005a,Nazarenko2006a,Jackson2009,Proment2009,Cockburn2010a,Wouters2010b,Wouters2010a, Sieberer2013a,Sieberer2013b,Tauber2013a,berloff2014}.
Superfluid turbulence can be formulated in a setting similar to the above Burgers turbulence problem, with the additional constraint that the forcing must conserve the property of the velocity to be a potential field.
In the SGPE, we allow for the necessary dissipation, loss, and gain of energy and particles by allowing $\mu=\mu_{1}+i\mu_{2}$ and $g=g_{1}-ig_{2}$ to become complex, including an effective particle gain or loss $\mu_{2}$, as well as two-body interaction and loss parameters $g_{1,2}$. 
The diffusion term $\propto \nu$ is generated through the coarse graining of high-frequency modes 
\cite{Wouters2010a,Wouters2010b, Sieberer2013a,Sieberer2013b,Tauber2013a}.
$\zeta$ is a Gaussian, delta-correlated white noise, i.e., $\langle\zeta^{*}(t,\mathbf{x})\zeta(t',\mathbf{x}')\rangle=\gamma\delta(t-t')\delta(\mathbf{x}-\mathbf{x}')$, induced by the loss and gain of particles.

Superfluid turbulence \cite{Vinen2006a,Maurer1998,Walmsley2007a,*Walmsley2008a,Nazarenko2001a,Araki2002a,Kobayashi2005a,Volovik2004a,Kozik2009a,Tsubota2008a,*Tsubota2010a,*Numasato2010a} manifests itself in self-similar field configurations in the domain of long-wavelength hydrodynamic excitations.
The hydrodynamic formulation of the SGPE results by introducing the parametrisation $\psi=\sqrt{n}\exp[\I\theta]$ in terms of the fluid density $n$ and velocity fields {$\mathbf{v}=m^{-1}\boldsymbol{\nabla}\theta$}.
The phase angle $\theta$ then obeys a Langevin equation of the KPZ type which is equivalent to Burger's equation \eq{Burgers} for the curl-free velocity field $\mathbf{v}$, under the  condition that {$\mathbf{f}=m^{-1}\boldsymbol{\nabla} U$}, with a random potential field $U$. 
See Appendix \app{superfluid_hydro} for details.

We note that the cubic non-linearity in the KPZ Hamiltonian can lead to an instability.
For typical parameter choices, however, the KPZ equation describes surface growth and smoothing  \cite{Kardar1986a}, and a steady state is reached when the driving and dissipation compensate each other.
Shocks in the velocity field, corresponding to cusps in the surface, develop and grow.
The dynamics described by the SGPE is different insofar the phase $\theta$ lives on the compact circle.
Moreover, the GPE supports solitary wave solutions and (quasi) topological defects such as vortices.
Velocity shock waves created due to the non-linearity typically lead to the formation of such defects.

%=======================================================================
%=======================================================================
\section{Renormalisation-group approach}
\label{sec:RGapproach}
The functional RG \cite{Wetterich:1992yh} provides a non-perturbative
framework to implement the coarse graining inherent to the RG. See
Refs.~\cite{Berges:2000ew,Polonyi:2001se,Pawlowski:2005xe,Delamotte:2007pf,Scherer:2010sv,Metzner:2011cw,Braun:2011pp,Berges2012a,Boettcher:2012cm} for  reviews.

%=======================================================================
\subsection{Wetterich's flow equation}
\label{sec:WsFlow}
A functional renormalisation-group (RG) analysis allows to determine the effective turbulent dynamics of the infrared (IR) modes by applying a Wilson-type averaging procedure to the ultraviolet {(UV)} modes.
This leads to an effective action $\Gamma_{k}[\mathbf{v}]$ describing the IR modes with $p\equiv|\mathbf{p}|<k$, by  integrating out the higher momenta,
\begin{align}
\text{e}^{-\Gamma_k[\mathbf{v}]} 
= \int \prod_{p>k\atop\omega} \, \text{d}\mathbf{v}(\omega,\mathbf{p}) \, \text{e}^{-S[\mathbf{v}]}. 
\label{eq:interpolation}
\end{align}
In the case of a Langevin field equation with a Gaussian forcing, the weight of the  field configurations can be expressed in terms of the exponential of the action
\begin{align}
S[\mathbf{v}] 
=  \frac{1}{2}\int_{t,\mathbf{x},\mathbf{y}} \!\!
{E_i(t,\mathbf{x}) \, E_j(t,\mathbf{y})} \, F_{ij}^{-1}(|\mathbf{x}-\mathbf{y}|) \, .
\label{eq:Burgersaction}
\end{align}
Here and in the following repeated indexes are to be summed over.
$\mathbf{E}(t,\mathbf{x})=\mathbf{E}(\mathbf{v}(t,\mathbf{x}))$ is defined as
\begin{align}
\mathbf{E}(t,\mathbf{x})=\partial_t {\bf v} + ({\bf v} \cdot \boldsymbol{\nabla}) {\bf v} - \nu \nabla^2 {\bf v}\, .
\end{align}
The expression in \Eq{Burgersaction} is obtained within the Martin-Siggia-Rose/Janssen-de Dominicis formalism \cite{Martin1973a,Bausch1976a,Janssen1976a,DeDominicis1978a,Zinnjustin2002a} by integrating out the response field. This field is usually explicitly kept in the action for having access to response functions which we do not consider here.
We implement the  cutoff $k$ by adding a term 
\begin{align}
  \Delta S_k[\mathbf{v}] 
  &=  \frac12\int_{\omega,\mathbf{p}} 
{v_{i}}(\omega,\mathbf{p}) R_{k,ij}(|\mathbf{p}|) {v_{j}}(-\omega,-\mathbf{p}) 
\end{align}
to the action $S$, choosing the regulator $R_{k,ij}(p)=\delta_{ij}R_k(p)$ diagonal in frequencies and momenta which diverges for $|\mathbf{p}|=p \ll k$, damping out the velocity fluctuations on scales larger than $1/k$.
For $p \gg k$, the regulator vanishes.
In the limit $k\to0$ the full effective action $\Gamma[\mathbf{v}]$ results, which takes into account all fluctuations and generates the physical correlation functions.
The regulator term allows to extend the functional integral over all momenta $p$ and define the coarse-grained Schwinger functional $W_k[\mathbf{J}]$,
\begin{align}
\text{e}^{-W_k[\mathbf{J}]} 
= \int \text{d}\mathbf{v} \, \text{e}^{-S[\mathbf{v}]- \Delta S_k[\mathbf{v}]+\int_{t,\mathbf{x}} \, \mathbf{J} \cdot \mathbf{v}}.
\label{eq:interpolation2}
\end{align}
{Functional derivatives of $W_k[\mathbf{J}]$ generate the coarse grained connected correlations functions.}
From this, the scale dependent effective action $\Gamma_k[v]$ is defined through the Legendre transform of $W_k[\mathbf{J}]$,
\begin{align}
{\Gamma}_k[v] = -W_k[\mathbf{J}] + \int_{t,\mathbf{x}} \mathbf{J} \cdot \mathbf{v}-\Delta S_k[\mathbf{v}].
\end{align}
It interpolates between the bare action \Eq{Burgersaction}, for $k\to \infty$, and the full effective action $\Gamma[\mathbf{v}]$, for $k\to 0$. The change of $\Gamma_k[\mathbf{v}]$ with the cutoff $k$ is  determined by Wetterich's flow equation \cite{Wetterich:1992yh}
\begin{align}
k \partial_k \Gamma_k[\mathbf{v}] 
= \frac{1}{2} \text{Tr}\left( \frac{1}{\Gamma_k^{(2)}[\mathbf{v}]+R_k}k\partial_k R_k \right)
\equiv I_{k}[\mathbf{v}].
\label{eq:wetterich}
\end{align}
This involves the second moment of the effective action, with matrix elements
$\Gamma_{k,ij}^{(2)}[\mathbf{v}](\omega',\mathbf{k}';\omega,\mathbf{k})=  {\delta^2 \Gamma_{k}}/({\delta v_{i}(\omega',\mathbf{k}')\delta v_{j}(\omega,\mathbf{k})})$. 
Evaluated on the solution of {$\Gamma_k^{(1)}[\mathbf{v}]=0$}, it is related to the inverse of the $k$-dependent two-point velocity cumulant,
\begin{align}
G_{k\to0,ij}(\omega,\mathbf{p})&=(2\pi)^{d+1} \langle v_{i}(\omega,\mathbf{p})v_{j}(-\omega,-\mathbf{p})\rangle
\nonumber\\
&=[{\Gamma_k^{(2)}}[0]+R_{k}]_{ij}^{-1}|_{k\to0}. 
\label{eq:Gk}
\end{align}
Details are deferred to Appendix \ref{sec:defgamma}.

%=======================================================================
\subsection{Approximation scheme}
\label{sec:approximation}
The RG flow equation \eq{wetterich} connects effective action functionals for different cutoff scales $k$  to each other.
Hence, in order to find self-similar turbulent configurations we look for IR fixed points of the flow, i.e., for solutions of \Eq{wetterich} which are scaling in $\omega$ and $p$ in the limit $k\to0$.
\Eq{wetterich} relates the action with its second moment and therefore creates an infinite hierarchy of integro-differential equations for its $n$th order moments.
It is a functional integro-differential equation. Its full solution is equivalent to solving the path integral. In the present case, we are interested in turbulent scaling solutions. Hence, the ansatz for the effective action should allow for these solutions and respect the Galilei symmetry of the underlying theory. We choose
\begin{align}
  \Gamma_k[\mathbf{v}]
  & = \frac{1}{2} \int_{\omega,\mathbf{p}} \mathbf{E}_k(\omega,\mathbf{p}) \cdot \mathbf{E}_k(-\omega,-\mathbf{p}) \, F^{-1}_k(p), 
  \nonumber \\
 \mathbf{E}_k (\omega,\mathbf{p}) 
 & = \left( i\omega + \nu_k(p) p^2\right) \mathbf{v}(\omega,\mathbf{p}) 
 \nonumber \\
 & \quad - i \int_{\omega',\mathbf{q}} 
 \left[\mathbf{v}(\omega-\omega',\mathbf{p}-\mathbf{q})\cdot \mathbf{q} \right] \, 
 \mathbf{v}(\omega',\mathbf{q})
\label{eq:trunc}
\end{align}
for the effective average action, in terms of the effective inverse force correlator $F_{k}^{-1}$ and the kinematic viscosity $\nu_{k}$.
$\Gamma_{k}[\mathbf{v}]$ has the same form as the bare action $S[\mathbf{v}]$ of the underlying Burgers equation \eq{Burgersaction}, but with the inverse force correlator and the kinematic viscosity allowed to be $k$-dependent.
We anticipate that $\nu_{k}(p)$ will become $p$-dependent as a result of the RG flow because the advective derivative renders the cubic and quartic couplings of the velocity field momentum dependent. \Eq{trunc} is the minimal ansatz consistent with Galilei symmetry and including the necessary momentum and frequency dependence for turbulence solutions. Extensions of the present approximation and the discussion of the systematic error will be presented elsewhere.

With the above ansatz, we keep a general dependence of the inverse propagator on $p$ while taking into account the $\omega$-dependence in an expansion to first order in $\omega^{2}$,
\begin{align}
\Gamma^{(2)}_{k,ij}(\omega,\mathbf{p}) 
& \equiv \delta_{ij} \, \Gamma^{(2)}_{k}(\omega,\mathbf{p}) \nonumber \\
& =  \delta_{ij} \left[ \nu_k(p)^2 p^4 +\omega^2\right]F^{-1}_k(p).
\label{eq:Gprop}
\end{align}
The sole dependence on the norm $p$ reflects the assumed rotational invariance.
The truncation of the frequency expansion ensures  that the integrand on the right hand side of the flow equation \eq{wetterich} is a rational function of $\omega$. 
See Appendix \ref{ap_I} for details.

Note that our ansatz is chosen such that the inverse force correlator $F_k^{-1}$ and thus the inverse propagator are diagonal in momentum $p$ and in the field indices $i,j$. 
As a consequence, the correlator describes field configurations where the forcing injects vorticity.
In contrast to this, Gaussian random forcing that keeps the velocity field curl free can be written as the gradient of a potential, 
$\mathbf{f} = \boldsymbol{\nabla}U$, and therefore obeys a noise correlator of the form
\begin{align}
\langle f_i(\omega,\mathbf{p}) f_j(\omega',\mathbf{p}') \rangle 
= \delta(\omega+\omega')\, \delta\left(\mathbf{p}+\mathbf{p'}\right) \, p_i \, p_j \, u(\omega,\mathbf{p}),
\label{eq:potential_forcing}
\end{align}
where $u$ is the scalar correlator of the potential.

%=======================================================================
\subsection{Flow equations for correlators}
\label{sec:FlowEqsCorr}
The truncated differential equations describing the RG flow of $F^{-1}_k(p)$ and $\nu_k(p)$ are obtained by taking field and frequency derivatives of \Eq{wetterich} and evaluating them at $\mathbf{v}=0$ and $\omega = 0$.
The diagonal ansatz \eq{Gprop} implies that the right-hand side of the resulting flow equations is projected onto the part diagonal in frequencies, momenta, and field indices, i.e.,
\begin{align}
& \frac{\delta^2 I_{k}}{\delta v_i(\omega,\mathbf{p}) \delta v_{j}(\omega',\mathbf{p}')} [{\mathbf{v}=0}] 
\equiv I_{k,ij}^{(2)}[0](\omega,\mathbf{p};\omega',\mathbf{p}')
\nonumber \\[0.3cm]
& \qquad \qquad 
\to \delta_{ij}
\frac{\delta(\omega+\omega')  \delta(\mathbf{p} + \mathbf{p}')}{(2\pi)^{d+1}} 
I_{k}^{(2)}[0](\omega,\mathbf{p}).
\label{eq:flow_integrals_first}
\end{align}
After the expansion \eq{Gprop} to order $\omega^{2}$ the flow equations formally read
\begin{align}
  k\partial_k \left[F^{-1}_k(p) \nu_k(p)^2 p^4 \right] 
 &= I_{k}^{(2)}[0](\omega = 0,p), \nonumber \\
  k\partial_k F^{-1}_k(p)\, 
 &= \frac{\partial}{\partial \omega^2} I_{k}^{(2)}[0](0,p).
\label{eq:flow1}
\end{align}
The second derivative of $I_{k}$ has the diagrammatic representation
\begin{equation}
 \begin{tikzpicture}[scale=0.6]
\draw[color=black, line width = 1.5pt] (1-0.4, 0) circle (1);
\draw[color=black, line width = 1.5pt] (4.7, 0) circle (1);
\path[draw] (-0.6*0.8/0.6-0.4,0) -- (0-0.4,0);
\path[draw] (2-0.4,0) -- (2+0.6*0.8/0.6-0.4,0);
\path[draw] (4.7-0.4243*0.8/0.6,-1-0.4243*0.8/0.6) -- (4.7,-1) -- (4.7+0.4243*0.8/0.6,-1-0.4243*0.8/0.6);
\node[draw,circle, fill=gray!30] (a) at (0-0.4, 0) {3};
\node[draw,circle, fill=gray!30] (b) at (2-0.4, 0) {3};
\node[draw,circle, fill=black] (c) at (1-0.4, 1) {};
\node[draw,circle, fill=gray!30] (d) at (4.7, -1) {4};
\node[draw,circle, fill=black] (e) at (4.7, 1) {};
\node (f) at (1.8+0.8/0.6, 0) {$ \displaystyle-\frac{1}{2}$};
\node (g) at (-1.9-1.2/0.6, 0) {$ \displaystyle  I_{k}^{(2)}[0](\omega,p)=$};
\end{tikzpicture}\, ,
\label{eq:flow_3}
\end{equation}
where the thick lines denote $G_{k}=(\Gamma_k^{(2)}+R_k)^{-1}$, the thin lines external momenta and frequencies, and the black dots insertions of the derivative $k\partial_{k}R_{k}$ of the regulator. The 3- and 4-vertices are given in the appendix, in \Eqs{Gamma3} and \eq{Gamma4}, respectively.

Since we have inserted the truncated effective action $\Gamma_k[{\bf v}]$ on both sides of the flow equations, the right hand sides of \Eqs{flow1} are again functionals of $\nu_k(p)$ and $F^{-1}_k(p)$ {which} are given explicitly in App.~\ref{ap_I},  Eqs.~\eq{explicit_flow_1} and \eq{explicit_flow_2}.
The resulting closed set of RG flow equations describes the change of $\nu_k(p)$ and $F^{-1}_k(p)$ under a shift of the cutoff scale $k$. 
For $k\to \infty$, all velocity fluctuations are suppressed, and we set (bare) initial conditions: $\nu_\infty(p) = \nu$, $F^{-1}_\infty(p) = F^{-1}(p)$.
In the infrared limit $\nu_{k\to0}(p)$ and $F^{-1}_{k\to0}(p)$ describe the {physically observable} viscosity and forcing as functions of $p$. 

%=======================================================================
%=======================================================================
\section{Fixed Points of the Renormalisation-group flow}
\label{sec:fixed_point_equations}
Universal scaling regimes are described by fixed points of the renormalisation-group (RG) flow: At an RG fixed point, the system becomes invariant under scaling transformations and exhibits scaling properties of correlations such as \eq{DeltavScaling}. In turn, fixed points encode the turbulent solutions of our model. In the following we apply the fixed point approach developed in \cite{Pawlowski:2003XX} for momentum and frequency dependent correlation functions.

%=======================================================================
\subsection{Parametrisation in the infrared}
As we approach the fixed points at vanishing cut-off, $k=0$, we {parametrise} the inverse propagator in terms of a scaling form.
Moreover, we make use of the rescaled variables, using units where the viscosity is dimensionless,
\begin{align}
& \hat{p} = \frac{p}{k}, & 
& \hat{\omega} = \frac{1}{k^2} \sqrt{\frac{z_2}{z_1}} \omega, & 
& \hat{\bf v}(\hat{\omega},\hat{\bf p}) = k^{d+1} {\bf v}(\omega,{\bf p}).
\label{eq:dim_variables}
\end{align}
Using these, the scaling forms of the propagator and its frequency derivative read \cite{Pawlowski:2003XX}:
\begin{alignat}{3}
 \Gamma^{(2)}_{k}(0,p) 
 & =  \, \Gamma^{(2)}_{0}(0,p) \left[1 + \delta Z_1\left(\hat{p}\right) \right], 
\nonumber \\[0.3cm]
 \left.{\partial_{\omega^{2}} \Gamma^{(2)}_{k}}\right|_{(\omega^2=0,p)} 
 & =  \left.{\partial_{\omega^{2}} \Gamma^{(2)}_{0}}\right|_{(\omega^2=0,p)}  \left[1 + \delta Z_2\left(\hat{p}\right) \right] \, ,
\label{eq:parametrization}
\end{alignat}
{where $ \delta Z_1 $ and $ \delta Z_2$ are the deviations of the two-point correlators from those at vanishing cutoff.}
Here and in the following, indices $i=1,2$ of $\delta Z_{i}$, etc., refer to the 0th and 1st-order derivatives with respect to $\omega^{2}$.
In the limit $\hat p\to\infty$, the conditions $\delta Z_i(\hat{p} \rightarrow \infty) = 0$ must be fulfilled.
Since this is the scaling limit, our propagator takes the form
\begin{alignat}{3}
& \Gamma^{(2)}_{0}(0,p) 
& & = k^{d} \, z_1 \, \hat{p}^{\eta_1}\,, 
\nonumber \\[0.3cm]
& \left.{\partial_{\omega^{2}} \Gamma^{(2)}_{0}}\right|_{(\omega^2=0,p)} 
& & = k^{d-4} \, z_2 \, \hat{p}^{\eta_2}\,,
\label{eq:Gamma20}
\end{alignat}
which, by definition, is independent of the scale $k$.
Hence,
\begin{align}
&z_1 \sim k^{\eta_1-d}, & 
& z_2 \sim k^{\eta_2-d+4}.
\label{eq:zi}
\end{align}
{Note that \Eqs{Gamma20} are used as definitions of $z_i$ and $\eta_i$. Then \Eqs{parametrization} define $\delta Z_i(p)$.
The powers of $k$ in front of \Eqs{parametrization}} indicate the dimensions of $\Gamma_{k}^{(2)}$ and its derivative.
The rescaled propagator, 
\begin{align}
\hat{\Gamma}^{(2)}_{k}(\hat{\omega},\hat{p}) = \Gamma^{(2)}_k(\omega,p) / (k^d z_1)\, ,
\end{align}
depends on $k$ only implicitly through $\hat p$. 
The fixed-point parametrisation \eq{parametrization} exhibits the IR scaling defined by the exponents $\eta_{i}$.
We find that the scale dependent functions $F^{-1}_k({p})$ and $\nu_k({p})$ are related to the variables introduced above by
\begin{alignat}{3}
& \nu_k(p) & & = \sqrt{z_1/z_2}\, \hat{p}^{(\eta_1-\eta_2-4)/2}  \sqrt{\frac{1+\delta Z_1(\hat{p})}{1+\delta Z_2(\hat{p})}}, 
\nonumber \\
& F^{-1}_k(p) & & = k^{d-4} z_2\, \hat{p}^{\eta_2} \left[1+\delta Z_2(\hat{p})\right].
\label{eq:nuOm_2_dZ}
\end{alignat}
The RG flow equations determine the possible values of the exponents $\eta_{1,2}$ at the fixed points of the RG flow.
At the fixed point, i.e., in situations where the parametrisation \eq{parametrization} is valid for all $k\ge0$, the functions $\delta Z_i(\hat{p})$ characterise the difference between coarse-grained effective actions $\Gamma_{k}$, which show scaling for $p\gg k$, and the (fully scale invariant) effective action obtained for $k\to0$. 
$\Gamma_{k \rightarrow 0}[{\bf v}] \equiv \Gamma[{\bf v}]$ does not depend on $k$ and generates physical observables. 
\Eq{DeltavScaling} implies that
\begin{align}
& \chi = \frac{\eta_1+\eta_2-2d}{4}+1, & z = \frac{\eta_1-\eta_2}{2}.
\label{eq:eta_to_chi}
\end{align}
%

%=======================================================================
\subsection{Fixed-point equations}
\label{sec:FP}
We obtain the fixed points by inserting \Eqs{nuOm_2_dZ} into the flow equations \eq{flow1} and expressing the resulting equations for the $\delta Z_{i}$ in terms of the  rescaled variables \eq{dim_variables},
\begin{align}
\frac{\text{d}\delta Z_i}{\text{d}\hat{p}} = -h \frac{\hat{I}^{(2)}_i(\hat{p})}{\hat{p}^{\eta_i+1}}{,\quad i=1,2.}
\label{eq:flow2}
\end{align}
Here, the rescaled flow integrals $\hat{I}^{(2)}_{i}(\hat{p})$ are related to the velocity-squared derivative of $I_k[\mathbf{v}]$ by
\begin{align}
& I_{k}^{(2)}[0]\left(\omega,p\right) 
= {k^d} \sqrt{\frac{z_2}{z_1}} \left[\hat{I}^{(2)}_1(\hat{p}) + \hat{\omega}^2 \hat{I}^{(2)}_2(\hat{p}) + \mathcal{O}\left(\hat{\omega}^4\right)\right]
\end{align}
and are given explicitly in \Eqs{I12hat}--\eq{rescaled_flow_2}, with \eq{F11}--\eq{F24}.
We note that the indices $1,2$ of $\hat{I}^{(2)}_{i}$ which refer to the order of the $\omega^{2}$ expansion replace the notation of the explicit cutoff dependence of $I^{(2)}_{k}$.
Before we discuss, in the next section, the properties of these flow integrals in further detail, we take a look at the prefactor
\begin{align}
h\equiv\sqrt{\frac{z_2}{z_1}} \frac{1}{z_1}
\label{eq:defh}
\end{align}
which equally appears in both flow equations.
$h$ must be independent of $k$ since all other terms in \Eqs{flow2} depend on $k$ only implicitly through $\hat p$.
$h$ is the effective coupling constant the theory assumes at the fixed-point.
A vanishing coupling $h=0$ implies that the fixed point is Gaussian.

The coupling's independence of $k$, together with the asymptotic scaling relations \eq{zi}, implies the relation
\begin{align}
 \eta_2+4+2d=3\eta_1
\label{eq:consistency}
\end{align}
between the  $\eta_{1,2}$.
Using this, we find the exponents  \eq{eta_to_chi} at the fixed point to be related to $\eta_{1}$ and $d$ only,
\begin{align}
 z &= 2+d-\eta_1,
 \label{eq:zofeta1}
 \\
 \chi&=\eta_{1}-d.
\end{align}
Adding these equations one obtains 
\begin{align}
 \chi+z = 2.
 \label{eq:chiofeta1andorz}
\end{align}
This can be attributed to Galilean invariance which prohibits an anomalous scaling of the velocity field \cite{Forster1977a,Medina1989a,Adzhemyan1999a,Kloss2012a,Berera2001a}:
Since the non-linearity of ${\bf E}$ is part of the advective derivative of the fluid velocity, it must scale in the same way as the partial time derivative. 
This is only possible if the velocity scales as position divided by time ($\sqrt{\Delta v} \sim r/r^z$), which implies, with the definition of $\chi$ and $z$ by \Eq{DeltavScaling}, the relation \eq{chiofeta1andorz}.
Of the four parameters $\eta_{1,2}$ and $z_{1,2}$ we have introduced only two are independent. 
Besides the relation \eq{consistency}, the overall scaling of the two-point function $G_k^{-1}(\omega,\mathbf{p})$ is not determined by the fixed-point equations and leaves a free choice of the system of units for $g$.

%=======================================================================
\subsection{Velocity correlations and kinetic energy}
\label{sec:observables}
Once the scaling exponents, $\eta_i$, are known, the scaling function, $g(x)$ can be derived.
This gives the scaling behaviour  \eq{DeltavScaling}  of the moment \eq{Deltav} which, with \eq{trunc}, reads
\begin{align}
& \Delta v_k (\tau,r)
= {d \int_{\bf p}}
\frac{1-\text{e}^{-\left|\tau\right| \nu_k(p)p^2} \text{e}^{-i {\bf p}\cdot {\bf r}}}{F^{-1}_k(p) \nu_k(p) p^2}.
\end{align}

Inserting \Eq{nuOm_2_dZ} and taking the limit $k \rightarrow 0$ one has $\delta Z_i(\hat{p})=0$ and obtains
\begin{align}
&\Delta v (\tau,r)
= {\frac{d}{ h} \frac{k^2}{z_1^2}}
\nonumber\\
&\times\int_{\bf \hat{p}}
\frac{1-\exp\left[- ({\left|\tau\right| k^2}/{z_1})({\hat{p}^{d+2-\eta_1}}/{h})-i \hat{\bf p}\cdot k{\bf r}\right]}{\hat{p}^{2\eta_1-2-d}}.
\end{align}
One can check, by substituting back ${\bf p}$ as integration variable, that the apparent dependence on the cutoff scale $k$ cancels out. 
We are free to choose $k = 1/r$ and write
\begin{alignat}{3}
\Delta v (\tau,r)
&&& = r^{2(\eta_1-d-1)}\left(\frac{k^{\eta_1-d}}{z_1}\right)^2\, 
\hat g\left(\frac{\tau}{r^{d+2-\eta_1}}\frac{k^{\eta_1-d}}{z_{1}}\right),
\nonumber \\[0.3cm]
\hat g(x)
&&&= {\frac{d}{h}  
\int_{\hat{\bf p}} \frac{1-\exp\left[{-{\hat{p}^{d+2-\eta_1}x}/{h }}{-i \hat{p}_z}\right] }{\hat{p}^{2\eta_1-2-d}},}
\label{eq:g}
\end{alignat}
where $\hat{p}_z$ is the $z$-component of $\hat{\bf p}$. 
According to the definition \eq{DeltavScaling} the above scaling form is consistent with the relations \eq{eta_to_chi}, \eq{zofeta1}--\eq{chiofeta1andorz} between  $\chi$, $z$, $\eta_{1}$, and $d$.
Moreover, the factor $k^{\eta_1-d}/z_1$, which makes the argument of $\hat g$ dimensionless, does not depend on the cutoff scale $k$, see \Eq{zi}, and is used to normalise the scaling function. See the discussion in \Sect{FP}.
Finally, inserting \Eq{Gprop} into \Eq{kin_energy_1} and performing the frequency integration, we obtain the momentum scaling of the kinetic energy density
\begin{align}
 \epsilon_{\text{kin}}({\mathbf{p}})
 &=\frac{1}{2} \int_{\omega} \, \langle \mathbf{v}(\omega,\mathbf{p}) \cdot \mathbf{v}(-\omega,-\mathbf{p})\rangle
 \nonumber\\
 &=
 \frac{d}{4F^{-1}_0\left(p\right) \nu_0\left(p\right) p^2} \sim p^{-2\eta_1+2+d}.
\label{eq:kin_energy}
\end{align}
Hence, according to \Eq{xi}, $\xi = 2\eta_1-2-d$.

%=======================================================================
\subsection{Non-Gaussian fixed points}
\label{sec:constraints}
In the remainder of this section we focus on non-Gaussian fixed points at which the coupling $h$ is non-vanishing, see \Eq{defh}.
We derive two further relations determining, together with the constraints discussed in the previous section, the four parameters $\eta_{1,2}$ and $z_{1,2}$.
This is possible without explicitly solving \Eqs{flow2}.
%The present subsection is, for most part, not relevant for \Sect{comparisons} and can be skipped at first reading.

The expressions of the flow integrals $\hat{I}^{(2)}_i(y)$ in \Eqs{flow2}, in terms of the functions $\delta Z_{i}$, are given in the appendix, in \Eqs{I12hat}--\eq{rescaled_flow_2}, with \eq{F11}--\eq{F24}.
To understand the relevance of the different contributions adding to these integrals let us consider in more detail both, the limits $\hat p\to0$ ($p\ll k$) and $\hat p\to\infty$ ($p\gg k$).

Any set of RG flow equations that is local in momentum scale must satisfy
\begin{align}
\lim_{p/k \to \infty} \frac{k\partial_k \Gamma_k^{(2)}(\omega,\mathbf{p})}{\Gamma_k^{(2)}(\omega,\mathbf{p})} =  0.
\label{eq:local_flow}
\end{align}
This means that, for $p$ far away from the cutoff $k$, the change of the effective action with $k$ does not affect the physics. 
We will see in the following that the flow equations at small $\hat{p}$ provide constraints on their solutions together with an explicit range of values for $\eta_1$. 
The opposite limit, $\hat{p} \gg 1$ allows to identify UV-divergent fixed points which can be related to a direct cascade of energy.

%=======================================================================
\subsubsection{Limit of momenta $p\ll k$}
\label{sec:hatpllone}
Before we discuss the physical limit where $p$ is much greater than the cutoff scale $k$, we consider the ``bare'' limit  of \Eqs{flow2} of small $\hat{p}$ where none of the fluctuations are integrated out.
We do this in order to be consistent with the order of the detailed derivations in \App{ScalingAnalysisFI} which is chosen such as to give simpler arguments first.
Even though we do not know the exact form of the effective action $\Gamma_{k\to \infty}[\mathbf{v}] \equiv \tilde{S}[\mathbf{v}]$ in this regime, we can extract information on the asymptotic scaling of $\Gamma_{k \to \infty}^{(2)}(\omega,\mathbf{p})$.
Indeed, all fluctuations being suppressed by the cutoff, it is natural to assume that the physical scaling of the propagator as given by $\eta_{1,2}$, cf.~\Eqs{parametrization} and \eq{Gamma20}, is absent. This is only possible if $\delta Z_i(\hat{p})$ assumes the asymptotic form
\begin{align}
\delta Z_i(\hat{p}\rightarrow 0) = -1 + a_i \hat{p}^{-\eta_i+\alpha_i}\left[1 + \mathcal{F}_{i}\left(\hat{p}\right) \right].
\label{eq:small_p}
\end{align}
Here, $a_i \hat{p}^{-\eta_i+\alpha_i}$ is the leading contribution to $\delta Z_{i}(\hat{p} \to 0)+1$, with constants $a_{i}$ and $\alpha_{i}$ to be determined, and $\mathcal{F}_{i}\left(\hat{p}\right)$ contains the sub-leading parts. One can check by inserting \Eq{small_p} into \Eq{parametrization} that
\begin{alignat}{3}
& \Gamma_{k \to \infty}^{(2)}(0,\mathbf{p}) & & \cong a_1 \, p^{\alpha_1}, \nonumber \\
& \left.{\partial_{\omega^{2}} \Gamma_{k \to \infty}^{(2)}}\right|_{(0,\mathbf{p})} & & \cong a_2 \, p^{\alpha_2}.
\label{eq:G_ktoinf}
\end{alignat}

Inserting \Eqs{small_p} into \Eqs{flow2} one finds that $\alpha_{i}-\eta_i-1$ are the exponents of the leading contributions to $\text{d}\delta Z_i/\text{d}\hat{p}(\hat{p}\to 0)$ on the left-hand sides. 
On the right-hand sides, the integrals fall into several terms, each with a different scaling behaviour in the limit $\hat p\to0$.
These terms are given explicitly in \Eqs{rescaled_flow_1}--\eq{F24} of \App{ScalingAnalysisFI}, their infrared scaling in \Eqs{IRscalingI1} and \eq{IRscalingI2}. 
The condition that the leading infrared scaling powers must be identical on both sides on \Eqs{flow2} leads to a closed set of equations for the exponents $\alpha_{i}$.
%=======================================================================
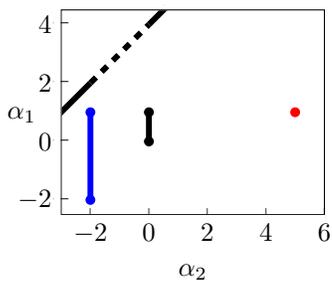
\begin{figure}[t]
\begin{tikzpicture}[scale=3.5]
\path[draw,dashed,line width=20/9pt] (-2/9,2/9) -- (0,4/9);
\path[draw,line width=20/9pt] (-31/90,1/10) -- (-2/9,2/9);
\path[draw,line width=20/9pt] (0,4/9) -- (1/15,23/45);
\draw[fill,white] (-3/9,4.5/9) rectangle (6/9,4.5/9+1/60);
\draw[fill,white] (-3/9-1/60,-2.5/9) rectangle (-3/9,4.5/9);
\draw (-3/9,-2.5/9) rectangle (6/9,4.5/9);
\draw[fill,color=blue] (-2/9,1/9) circle (0.15/9);
\path[draw,color=blue, line width=20/9pt] (-2/9,1/9) -- (-2/9,-2/9);
\draw[fill,color=red] (5/9,1/9) circle (0.15/9);
\draw[fill,color=blue] (-2/9,-2/9) circle (0.15/9);
\draw[fill,color=black] (0,0) circle (0.15/9);
\draw[fill,color=black] (0,1/9) circle (0.15/9);
\path[draw,color=black, line width=20/9pt] (0,0) -- (0,1/9);
%\path[] (-2/9,-2.5/9) node [below] {$-2$} -- (1.5/9,-2.5/9) node [below] {\begin{tabular}{c}  \\ $\alpha_2$ \end{tabular}} -- (0,-2.5/9) node [below] {$0$} -- (2/9,-2.5/9) node [below] {$2$} -- (4/9,-2.5/9) node [below] {$4$} -- (6/9,-2.5/9) node [below] {$6$};
%\path[] (-1/3,-2/9) node [left] {$-2$} -- (-1/3,0) node [left] {$0$} -- (-1/3,1/9) node [left] {\begin{tabular}{cc} $\alpha_1$ & \end{tabular}} -- (-1/3,2/9) node [left] {$2$} -- (-1/3,4/9) node [left] {$4$};
\path[] (-2/9,-2.5/9) node [below] {$-2$} -- (-2/9,-2.5/9) node {\tiny $|$} -- (1.5/9,-2.5/9) node [below] {\begin{tabular}{c}  \\ $\alpha_2$ \end{tabular}} -- (0,-2.5/9) node [below] {$0$} -- (0,-2.5/9) node {\tiny $|$} -- (2/9,-2.5/9) node [below] {$2$} -- (2/9,-2.5/9) node {\tiny $|$} -- (4/9,-2.5/9) node [below] {$4$} -- (4/9,-2.5/9) node {\tiny $|$} -- (6/9,-2.5/9) node [below] {$6$} -- (6/9,-2.5/9) node {\tiny $|$};
\path[] (-1/3,-2/9) node [left] {$-2$} -- (-1/3,-2/9) node {\bf --} -- (-1/3,0) node [left] {$0$} -- (-1/3,0) node {\bf --} -- (-1/3,1/9) node [left] {\begin{tabular}{cc} $\alpha_1$ & \end{tabular}} -- (-1/3,2/9) node [left] {$2$} -- (-1/3,2/9) node {\bf --} -- (-1/3,4/9) node [left] {$4$} -- (-1/3,4/9) node {\bf --};
\draw[fill,white] (-3/9,-2.53/9) rectangle (6/9,-2.75/9);
\draw[fill,white] (-3.04/9,-2.5/9) rectangle (-3.24/9,4.5/9);
\end{tikzpicture}
\caption{(Color online) Non-Gaussian fixed points: 
A graphical representation of the values of the bare exponents $\alpha_1$ and $\alpha_2$, defined in \Eq{small_p} in blue and red for $d=1$ and black and red for $d \neq 1$, which characterise the non-Gaussian fixed points.
The scaling exponents of the bare stochastic Burgers equation, as described by $S[{\bf v}]$, \Eq{Burgersaction}, are related by $\alpha_1 = \alpha_2 +4$ and are shown as a black line. 
The blue dot at $(\alpha_1,\alpha_2) = (1,-2)$ corresponds to the fixed point investigated in \cite{Kloss2012a,Medina1989a}, while the top half of the blue line (for $\alpha_1 >0$) represents the set of points found in \cite{Medina1989a} for different types of forcing, corresponding to a bare action with exponents given by the dotted black line, see the end of section \sect{hatpggone}.
In all dimensions we find a continuum of fixed points, and a further, new fixed point that may arise if the fluid is forced on large scales (red dot). Along the blue line we find that the scaling exponent satisfies $\eta_1 = 2-\alpha_1/2$ while it is $\eta_1 = 2-\alpha_1/2+d$ along the black line. $\eta_2$ is related to $\eta_1$ through \Eq{consistency}.
}
\label{fig:alphas}
\end{figure}
%=======================================================================

Due to the absence of angular integrations the case $d=1$ is special.
In $d=1$ dimension, the resulting constraints on the $\alpha_i$ read
\begin{align}
\alpha_1
&  =  \text{min}\left(\alpha_1+\alpha_2+2,\frac{\alpha_1}{2}+\frac{\alpha_2}{2}+2,1\right),
 \nonumber \\
\alpha_2
& = \text{min}\left(2\alpha_2,\alpha_1+\alpha_2,\frac{\alpha_1}{2}+\frac{\alpha_2}{2},\alpha_2\right)+2.
\label{eq:eq_alphas}
\end{align}
This restricts the values of $\alpha_{1,2}$ to the combinations
\begin{align}
&\left(\alpha_1,\alpha_2\right) \in \left(\left[-2,1\right],-2\right), \nonumber \\
&\left(\alpha_1,\alpha_2\right) = \left(1,5\right).
\label{eq:alphas}
\end{align}
In $d\neq 1$ dimensions, the leading IR scaling of the flow integrals is modified, but the same procedure leads to constraints on the $\alpha_{i}$:
\begin{align}
\alpha_1
& = \text{min}\left(1,2+\frac{\alpha_1}{2}+\frac{\alpha_2}{2},\alpha_1+\alpha_2\right), \nonumber \\
\alpha_2
& = \text{min}\left(2+\alpha_2,2+\frac{\alpha_1}{2}+\frac{\alpha_2}{2},\alpha_1+\alpha_2,2\alpha_2\right).
\label{eq:eq_alphasdD}
\end{align}
Solutions of these equations are the combinations
\begin{align}
&\left(\alpha_1,\alpha_2\right) \in \left(\left[0,1\right],0\right), \nonumber \\
&\left(\alpha_1,\alpha_2\right) = \left(1,5\right).
\label{eq:alphasdD}
\end{align}
The above results suggest that the allowed combinations ($\alpha_1,\alpha_{2})$, summarised in \Fig{alphas}, correspond to the different possible non-Gaussian fixed points. 
For the different dimensions, we find, for the scaling exponents relevant in the limit $p\ll k$, a connected interval for $\alpha_{1}$ and an additional point at $\left(\alpha_1,\alpha_2\right) = \left(1,5\right)$.
We find that none of the resulting small-$\hat{p}$ scalings of $\tilde{S}[{\bf v}]$ corresponds to that of the bare action $S[{\bf v}]$ for Burgers' equation.
In the bare action, $\nu$ is $p$-independent, which implies that the ratio of {$\Gamma_{k \to \infty}^{(2)}(0,p)$ and $\partial \Gamma_{k \to \infty}^{(2)}/\partial \omega^2 (0,p)$} scales as $p^4$, see \Eq{Gprop}. 
Hence, taking the ratio of \Eqs{G_ktoinf} gives $\alpha_1 = \alpha_2 +4$. 
Analogously, \Eq{forcingcorr} implies that {$\partial \Gamma_{k \to \infty}^{(2)}/\partial \omega^2 (0,p)$} scales as $p^{-\beta}$ and thus that $\alpha_2 = - \beta$.
The resulting possible combinations $(\alpha_{1},\alpha_{2})=(4-\beta,-\beta)$ are marked by the black (dashed/solid) line in \Fig{alphas}.
{As} expected, there is no choice of the forcing exponent $\beta$ that makes the stochastic Burgers equation sit at a non-Gaussian RG fixed point for all values of $k$.

%=======================================================================
\subsubsection{Limit of momenta $p\gg k$}
\label{sec:hatpggone}
In the opposite limit of vanishing cutoff all fluctuations are integrated out and the full effective theory emerges. 
Inserting Eqs.~\eq{parametrization}, the local-flow requirement \eq{local_flow} reads
\begin{align}
 \lim_{\hat{p} \to \infty} \frac{\hat{p}\,\delta Z_i'(\hat{p})}{1+\delta Z_i(\hat{p})} = 0,
\end{align}
with the notation $\delta Z_i'(\hat{p}) = \text{d} \delta Z_i(\hat{p})/\text{d}\hat{p}$.
Assuming that $\delta Z_i(\hat{p}\gg1)$ behaves as a power law, one finds that the above requirement is only fulfilled if $\lim_{\hat{p}\to\infty} \delta Z_i(\hat{p}) = 0$.
In this case, once the cutoff scale is sent to zero, see \Eq{parametrization}, one finds fixed points with correlations given by scaling functions across all momentum scales.
If, on the other hand, $\lim_{\hat{p}\to\infty} \delta Z_i(\hat{p}) = \infty$, an RG fixed point can only exist if the scaling range is restricted to momenta smaller than some upper cutoff $\Lambda$. 
Then scaling only arises within the range of physical momenta and $\delta Z_i(1\ll \hat{p}< \Lambda/k) \cong 0$. 
In this situation of a UV-divergent fixed point, the theory is not well defined exactly at the fixed point but the latter can be approached arbitrarily by choosing $\Lambda$ accordingly large.

Here, we consider UV-finite fixed points and take the limit $\hat{p}\to \infty$ in the flow integrals. 
The boundary condition $\delta Z_i(\hat{p}\rightarrow \infty)=0$ then allows us to write \Eq{flow2} in the integral form 
\begin{align}
& \delta Z_i\left(\hat{p}\right)
= h \int_{\hat{p}}^\infty \text{d}y \frac{\hat{I}^{(2)}_i(y)}{y^{\eta_i+1}}, \quad i=1,2.
\label{eq:flow3}
\end{align}
For $\hat{p}\gg 1$, also the integration variable $y$ exceeds $1$ by far such that we can approximate $\delta Z_i(y)=0$ in the integrals $\hat{I}^{(2)}_i(y)$, see Appendix \app{ScalingAnalysisFI} for a detailed discussion. 
The flow integrals can be further approximated by keeping only their leading term as $\hat{p} \to \infty$, 
\begin{align}
\hat{I}^{(2)}_i(\hat{p}\to\infty) \sim \hat{p}^{\beta_{i}}.
\label{eq:large_p}
\end{align}
We determine the exponents $\beta_{i}$ in \App{ScalingAnalysisFI}, see Eqs.~\eq{beta1} and \eq{beta2}.
In order to obtain finite integrals on the right hand side of Eqs.~\eq{flow3}, it is necessary that $\beta_i < \eta_i$. 
This implies that $\eta_{1}$ has to be within the range
\begin{align}
\left. \begin{array}{cc} 
d=1: & d \\ 
d\neq 1: & ({d+4})/{3} \end{array} \right\} < \eta_1 < d+1
\label{eq:range}
\end{align}
%
%=======================================================================
\begin{figure}[t]

\begin{tikzpicture}[scale=1.7]
\path[fill,color=gray!50] (1/2,3/2) -- (5/2,7/2) -- (1/2,7/2) -- cycle;
\path[fill,color=gray!50] (1/2,3/2) -- (7/2,5/2) -- (7/2,3/4) -- (1/2,3/4) -- cycle;
\path[fill,color=black] (1/2,5/2) -- (3/2,7/2) -- (1/2,7/2) -- cycle;
\path[draw,color=white, line width=3pt] (1,2) -- (1,1);
\path[draw,color=gray!120, line width=3pt] (1,3/2) -- (1,2.8);
\path[fill,color=gray!120, line width=2pt] (1/2,2) -- (2,7/2) -- (3/2,7/2) -- (1/2,5/2) -- cycle;
\path[draw,color=blue, line width=1pt] (1,3/2) -- (1,2);
\draw[fill,color=blue] (1,3/2) circle (0.05);
\draw[fill,color=blue] (2,2.379) circle (0.05);
\draw[fill,color=blue] (3,3.300) circle (0.05);
\draw[dashed,line width=1pt] (1/2,3/2) -- (5/2,7/2);
\draw[dashed,line width=1pt] (1/2,2) -- (2,7/2);
\draw[fill,color=white] (1/2,1/2) rectangle (1/2-1/40,7/2);
\draw[fill,color=white] (1/2,7/2) rectangle (7/2+2/40,7/2+2/40);
\draw[fill,color=white] (7/2,3/4) rectangle (7/2+2/40,7/2+2/120);
%\path[] (1,3/4) node [below] {$1$} -- (2,3/4) node [below,text width=1cm,align=center] {$2$ \\ $d$} -- (3,3/4) node [below] {$3$};
%\path[] (1/2,1) node [left] {$1$} -- (1/2,2) [left] node {$\eta_1$ $2$} -- (1/2,3) [left] node {$3$};
\path[] (1,3/4) node [below] {$1$} -- (1,3/4) node {\tiny $|$} -- (2,3/4) node [below,text width=1cm,align=center] {$2$ \\ $d$} -- (2,3/4) node {\tiny $|$} -- (3,3/4) node [below] {$3$} -- (3,3/4) node {\tiny $|$};
\path[] (1/2,1) node [left] {$1$} -- (1/2,1) node {\bf --} -- (1/2,2) node [left] {$\eta_1$ $2$} -- (1/2,2) node {\textcolor{white}{\bf --}} -- (1/2,3) node [left] {$3$} -- (1/2,3) node {\textcolor{white}{\bf --}};
\draw[fill,white] (1/2,3/4) rectangle (7/2,2.8/4);
\draw[fill,white] (1/2,3/4) rectangle (0.85/2,7/2);
\end{tikzpicture}

\caption{(Color online) The range of values of $\eta_{1}$ (defined {in \Eqs{Gamma20}} and related to $\eta_{2}$ in \Eq{consistency} and to $\chi$ and $\eta$ in \Eqs{eta_to_chi}), which correspond to UV-convergent non-Gaussian fixed-points for different spatial dimensions $d$ (white area).
The stripe at $d=1$ only applies to this single dimension.
In the regions shaded in light and dark grey any potential fixed point is UV-divergent such that the RG flow integrals must be regularised in the UV.
The blue dots correspond to the average literature values for the exponent $\eta_{1}$, cf.~Refs.~\cite{Tang1992a,Ala-Nissila1993a,Castellano1999a,Marinari2000a,Aarao2004a,Ghaisas2006a,Kelling2011a}.
Their values, as given in Table I. of Ref.~\cite{Kloss2012a}, are $\eta_{1}=1.5$ ($d=1$),  $\eta_{1}=2.379(15)$ ($d=2$), and $\eta_{1}=3.300(12)$ ($d=3$). 
The vertical blue line marks the range of possible exponents found in Ref.~\cite{Medina1989a} for different forcings, with the exponent $\beta$, defined in \Eq{forcingcorr}, chosen between $0$ and $2$. 
The dark-grey shaded area marks the range of $\eta_1$ corresponding to the sets $(\alpha_{1},\alpha_{2})$ shown as the blue and black solid lines at $\alpha_{2}=-2$ and $0$ in \Fig{alphas}, respectively. 
It continues accordingly also for $\eta_{1}>7/2$.
In between the two dotted lines there is a direct cascade of kinetic energy. 
In the black area at the left top, the dynamical critical exponent $z$ is negative.
See~\Sect{cascades} for further details.
}
\label{fig:eta_range}
\end{figure}
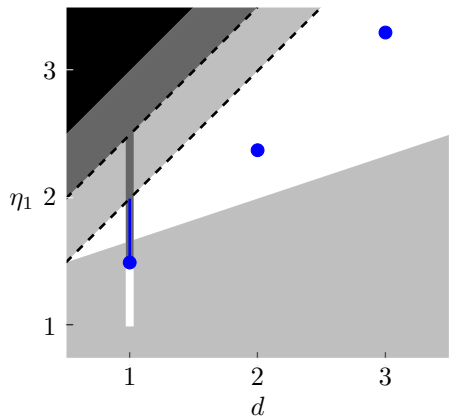
%=======================================================================
which is shown as the white area in \Fig{eta_range}.

%=======================================================================
\subsubsection{Implications for driving, and turbulent cascades}
\label{sec:cascades}
The bounds \eq{range} on $\eta_1$ have distinct physical interpretations: 
While the lower bound can be expressed as a regularity condition on the type of forcing that is sampled by the stochastic process, the upper bound marks the onset of a direct energy cascade.
 
As the forcing is a Gaussian random variable, it follows the probability distribution
\begin{align}
 P_{k}[\mathbf{f}] \sim \exp\left\{-\frac{1}{2}\int_{\omega,\mathbf{p}} \left|\mathbf{f}(\omega,\mathbf{p})\right|^2 F_k^{-1}(p)\right\}.
\end{align}
This distribution implies that the probability of a spatially local force field $\mathbf{f}(t,\mathbf{x}) \sim \delta(\mathbf{x}-\mathbf{x}_{f})$ can be finite if $\int_{\mathbf{p}} F_k^{-1}(p)$ is finite and is necessarily zero otherwise.
Furthermore, for the latter integral to be finite, the critical exponent at a  UV-finite fixed point needs to fulfil $\eta_1<(d+4)/3$, as one finds by inserting the parametrisation \Eq{nuOm_2_dZ} for $F_{k}^{-1}$. 
We conclude that in the lower grey shaded area of \Fig{eta_range},  $\eta_1 < (d+4)/3$, local Gaussian forcing $\mathbf{f}(t,\mathbf{x})$ is included at a UV-finite fixed point while it is suppressed for $\eta_1 > (d+4)/3$.
We have shown above that UV-finite non-Gaussian fixed points require $\eta_1 > (d+4)/3$.
Hence,  for an RG fixed point to be UV-finite, the forcing needs to be sufficiently regular in space-time, specifically {$\lim_{p\to\infty}\left|f(\omega,\mathbf{p})\right| = \lim_{\omega\to\infty}\partial_\omega\left|f(\omega,\mathbf{p})\right|=0$}. 

Note that, in the case $d=1$, the regularity condition is modified. 
The fixed points are UV-finite above the relatively lower limit $\eta_1 > 1$.
This is a consequence of the fact that in $d=1$ dimension, point-like shocks are stable solutions. 
Applying a force such as $\mathbf{f}(t,\mathbf{x}) \sim \delta(\mathbf{x}-\mathbf{x}_{s})$ is comparable to inserting a shock at the position $\mathbf{x}_{s}$.

The upper bound, $\eta_1 = d+1$, can be related to the appearance of a direct cascade of energy.
We briefly sketch the argument leading to this result in the following but leave a detailed derivation to a forthcoming publication.
In order for the system to be stationary, the equation of motion \eq{Burgers}, multiplied by $\mathbf{v}$ and averaged over statistically,  gives an evolution equation for the mean kinetic energy,
\begin{align}
\partial_t \frac{1}{2}& \langle v^2(t,\mathbf{x}) \rangle 
 = \nu \langle \mathbf{v}(t,\mathbf{x}) \cdot \Delta \mathbf{v}(t,\mathbf{x}) \rangle 
+ \langle \mathbf{f}(t,\mathbf{x}) \cdot \mathbf{v}(t,\mathbf{x}) \rangle 
\nonumber \\
& - \langle \mathbf{v}(t,\mathbf{x}) \cdot \left[ \left( \mathbf{v}(t,\mathbf{x}) \cdot \boldsymbol{\nabla}\right) \mathbf{v}(t,\mathbf{x})\right] \rangle = 0.
\label{eq:energy_balance}
\end{align}
The third term on the right-hand side can be written as
\begin{align}
\epsilon_\mathrm{adv}=\frac{1}{2} \langle \left( \mathbf{v}(t,\mathbf{x}) \cdot \boldsymbol{\nabla}\right) v^2(t,\mathbf{x}) \rangle
\label{eq:energy_transport}
\end{align}
and describes the advective increase of compressible energy at $\mathbf{x}$. 
Furthermore, 
\begin{align}
& \epsilon_{\nu} 
= - \nu \langle \mathbf{v}(t,\mathbf{x}) \cdot \Delta \mathbf{v}(t,\mathbf{x}) \rangle 
= \int_{\omega,\mathbf{p}} \nu_k(p)p^2 \langle \left|\mathbf{v}(\omega,\mathbf{p})\right|^2\rangle ,
\nonumber\\
& \epsilon_{f} 
= \langle \mathbf{f}(t,\mathbf{x}) \cdot \mathbf{v}(t,\mathbf{x}) \rangle
\label{eq:energy_dissipation_rate}
\end{align}
capture energy dissipation and injection rates, respectively.
Inserting the truncation \eq{trunc} into the energy dissipation rate, the frequency integration can be performed by means of the residue theorem, and one obtains, in the limit of removed regulator,
\begin{align}
 \epsilon_{\nu} 
 = \frac{d}{2}\int_{\mathbf{p}} F_{k=0}(p) 
 \equiv \int_0^\infty \text{d}p \, \epsilon_{\nu}(p).
\end{align}
The cubic advective term involves two convolutions in momentum space,
\begin{align}
 \epsilon_{\text{adv}} = \int_0^\infty \text{d}p \int_0^\infty \text{d}q \, \epsilon_{\text{adv}}(p,q).
\end{align}
The advective transport kernel $\epsilon_{\text{adv}}(p,q)$ describes the rate of kinetic energy that is transported from the momentum $q$ to  $p$. 
By means of the ansatz \eq{trunc}, $\epsilon_{\text{adv}}(p,q)$ can be expressed in terms of $\nu_k(p)$ and $F_k^{-1}(p)$.
In the limit $k\to0$ it then is a function of $h$ and $\eta_1$ only. 
Evaluating $\epsilon_{\text{adv}}(p,q)$ in this way, a direct energy cascade, i.e., transport which is local in momentum space on a logarithmic scale, can be identified for $d+1<\eta_1<d+3/2$.
In this regime, $\epsilon(p,q)$ is non-vanishing only for $p\cong q$ (locality), positive for $p<q$ and negative for $p>q$ (positive directionality), and $\epsilon(p,q)\simeq-\epsilon(p,-q)$ (balance of driving and dissipation, i.e., inertial turbulent transport). 

Note that it is natural to have a direct cascade requiring a UV regulator:
Physically, a cascade is realised only in a given inertial range. 
For example, in turbulence of an incompressible fluid in three dimensions, energy is injected on the largest scales and transported to smaller scales by the non-linear dynamics which leads to larger eddies feeding into smaller ones. 
The kinetic energy is dissipated into heat once it reaches the end of the inertial range set by the viscosity. 
At an RG fixed point, the inertial range by definition extends over all momenta.
Hence, the UV cutoff of the dissipation scale is absent.
As a result, energy in a direct cascade is transported to infinitely large momenta, leading to a UV divergence of the fixed-point theory.

Finally, using the solutions of Eqs.~\eq{eq_alphas} and \eq{eq_alphasdD}, the monomials in the flow integrals dominating in the limit $\hat{p}\to0$ can be identified, as is discussed in detail in Appendix \app{ai}, cf.~Eqs.~\eq{IRscalingI1} and \eq{IRscalingI2}. 
Equations for the $a_{i}$ are obtained by matching the corresponding pre-factors. 
Other than Eqs.~\eq{eq_alphas} and \eq{eq_alphasdD} these are not closed and can not be solved independently of \Eq{flow2}. 
However, for the interval sets of $\alpha_{i}$ marked by the black and blue lines in \Fig{alphas},
the ratio of \Eqs{aclosed_1d} (\Eqs{aclosed_Dd} for $d\neq1$) determining the pre-factors $a_{i}$ give new equations \eq{ratioaiEqs1d} and \eq{ratioaiEqsDd}, respectively, which are independent of the $a_{i}$ and provide the following relations between $\alpha_1$ and $\eta_1$,
\begin{align}
\eta_1 = \left\{\begin{array}{ll}
2-\alpha_1/2, & d=1\\
2-\alpha_1/2+d, & d\neq 1
\end{array}\right. .
\label{eq:eta12alpha1}
\end{align}
Specifically, for $d = 1$, we get $3/2<\eta_1<3$ and, for $d \neq 1$, $3/2+d<\eta_1<2+d$, intervals which we mark by dark grey shading in \Fig{eta_range}. 
This regime, for $d\not=1$, is disjunct with the regime where a direct energy cascade may occur.
Only for $d=1$, it allows for such a cascade, but does not require it.
In view of Refs.~\cite{Janssen1999a,Kloss2013a}, we note that we find an upper bound to the regime of allowed $\eta_{1}$.
Specifically, \Eq{zofeta1} implies that above this bound, the dynamical critical exponent would become negative.

%=======================================================================
%=======================================================================
\section{Implications for classical and quantum turbulence}
\label{sec:comparisons}
%
%=======================================================================
\subsection{Burgers and KPZ scaling  in $d=1$}
\label{sec:results1D}
\begin{figure}[t]

\begin{tikzpicture}[scale=4]
\path[draw,color=blue,line width=20/9pt] (-1/10,-2/5) -- (1/2,0) -- (2/3,0);
\draw[fill,color=gray!30] (-1/3,-1.25/3) rectangle (2/3,-1/3);
\draw[fill,color=gray!30] (-1/3,1/3) rectangle (2/3,1.25/3);
\draw (-1/3,-1/3) rectangle (2/3,1/3);
%\path[] (-1/3, -1.25/3) node [below] {$-1$} -- (0,-1.25/3) node [below] {$0$} -- (00.5/3,-1.25/3) node [below] {\begin{tabular}{c} \\ $\beta$\end{tabular}} -- (1/3,-1.25/3) node [below] {$1$} -- (2/3,-1.25/3) node [below] {$2$};
%\path[] (-1/3,-1/3) node [left] {$-1$} -- (-1/3,0) node [left] {\begin{tabular}{cc} $-\xi$ & $0$\end{tabular}} -- (-1/3,1/3) node [left] {$1$};
\path[] (-1/3, -1.25/3) node [below] {$-1$} -- (-1/3, -1.25/3) node {\tiny $|$} -- (0,-1.25/3) node [below] {$0$} -- (0,-1.25/3) node {\tiny $|$} -- (00.5/3,-1.25/3) node [below] {\begin{tabular}{c} \\ $\beta$\end{tabular}} -- (1/3,-1.25/3) node [below] {$1$} -- (1/3,-1.25/3) node {\tiny $|$} -- (2/3,-1.25/3) node [below] {$2$} -- (2/3,-1.25/3) node {\tiny $|$};
\path[] (-1/3,-1/3) node [left] {$-1$} -- (-1/3,0) node [left] {\begin{tabular}{cc} $-\xi$ & $0$\end{tabular}} -- (-1/3,0) node {\bf --} -- (-1/3,1/3) node [left] {$1$};
\path[] (0.3/3,0.7/3) node {$\leftarrow$ Non-local forcing} -- (1/3,0.3/3) node {Local forcing $\rightarrow$};
\draw[fill,white] (-1/3,-1.25/3) rectangle (2/3,-1.32/3);
\draw[fill,white] (-1.01/3,-1.25/3) rectangle (-1.1/3,1.25/3);
\end{tikzpicture}

\caption{(Color online) The scaling exponent of $\epsilon_{\text{kin}}(p) \sim p^{-\xi}$ for $d=1$, for different values of $\beta$. 
The grey area corresponds to values excluded for {UV-finite fixed points}.
The linearly rising part corresponds to the values of $\eta_1$ associated with the blue line of \Fig{alphas}, and the saturation value to the upper blue dot, cf.~\Eq{medinasol}.}
\label{fig:1d_scaling}
\end{figure}
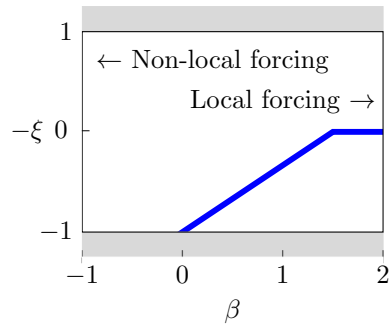

Our approach gives us access to the full set of fixed points that are consistent with the chosen truncation. 
In the following we discuss results known in the literature for Burgers and KPZ scaling solutions \cite{Medina1989a,Kloss2012a} in view of the conditions \eq{alphas},  \eq{range}, and \eq{eta12alpha1}.
$\eta_1$ has been related, within 1-loop perturbation theory \cite{Medina1989a}, to the exponent $\beta$ of the force correlation function \eq{forcingcorr}, for $d=1$ and $0<\beta\leq2$: 
\begin{align}
\eta_1 = \text{max}\left(\frac{3}{2},-\frac{\beta}{3}+2\right).
\label{eq:medinasol}
\end{align}
Hence, $\eta_{1}=2$ for $\beta = 0$, from where it decreases linearly as $\beta$ increases and saturates at $\eta_{1}=3/2$ for $\beta\ge 3/2$.
The kinetic energy spectrum scales as $\epsilon_{\text{kin}}\sim p^{-1}$ for a moderately local forcing ($\beta = 0$), and its exponent grows with $\beta$, i.e., $\epsilon_{\text{kin}}\sim p^{0}$ for $\beta \ge 3/2$, see \Fig{1d_scaling}.

{It was found numerically for $d=1$ and different values of $\beta$ \cite{Hayot1996a,Mesterhazy2013b} and for $d\geq 1$ \cite{Tang1992a,Ala-Nissila1993a,Castellano1999a,Marinari2000a,Aarao2004a,Ghaisas2006a,Kelling2011a} that the driven-dissipative KPZ dynamics reaches a steady state at large times where correlation functions exhibit scaling behaviour as in \Eq{DeltavScaling}.
The picture that emerges from these numerical investigations is consistent with the analytical results cited above.}

%=======================================================================
%=======================================================================
%

While the  combinations $(\alpha_{1},\alpha_{2})$, derived  in the previous section, cf.~\Eq{alphas}, allow us to predict the existence of new non-Gaussian RG fixed points, they are also fully consistent with the known perturbative results \cite{Medina1989a}.
Relation~\eq{eta12alpha1}, which applies on the solid blue line $-2<\alpha_1<1$ in \Fig{alphas}, includes the range $3/2\le \eta_1\le 2$ outside the shaded area in \Fig{eta_range} where the flow integral is UV-convergent (solid blue line in \Fig{eta_range}).
In the perturbative calculation of Ref.~\cite{Medina1989a}, the forcing exponent $\beta$ of the bare action is considered as a fixed parameter and governs the existence and properties of the different fixed points.
The resulting bare actions correspond to the set of $(\alpha_{1},\alpha_{2})$ marked as a dashed line in \Fig{alphas}.
It is consistent with the constraint $\alpha_{1}=\alpha_{2}+4$ resulting from the condition that the bare viscosity is $p$-independent, see the discussion after \Eq{alphasdD}.
In contrast to these actions, the set of $(\alpha_{1},\alpha_{2})$ marked as a blue line in the figure represents bare actions which remain at the fixed point throughout the RG flow.

When the forcing is strongly non-local, i.e., $\beta<0$, additional non-linearities become relevant and the loop expansion of Ref.~\cite{Medina1989a} fails. Our results show that negative values of $\beta$ can nevertheless be considered and that there is another, yet unobserved fixed point, $\left(\alpha_1,\alpha_2\right) = \left(1,5\right)$, with new critical exponents still to be determined and that this fixed point will be realised when $\beta<0$ and $h>0$ since it has a large positive value of $\alpha_2$.

For $d=2, 3$, we find that the literature values for $\eta_{1}$, see Refs.~\cite{Tang1992a,Ala-Nissila1993a,Castellano1999a,Marinari2000a,Aarao2004a,Ghaisas2006a,Kelling2011a}, and their averages quoted in Table I. of Ref.~\cite{Kloss2012a} and marked as blue dots in \Fig{eta_range}, are within the white range  defined in \eq{range}. 
They are hence outside the dark grey area in \Fig{eta_range} which we ascribe to the driving of velocity fields of non-vanishing curl.
We recall that, however, the Burgers and KPZ equations are equivalent only for curl-free fields, which in turn is not ensured by the truncation \eq{trunc} of the action as it involves a diagonal forcing correlator $F_{ij}\sim\delta_{ij}F$, see the discussion at the end of \Sect{approximation}.

%=======================================================================
\subsection{Strong wave and quantum turbulence}
\label{sec:resultsSWT}
In the following we discuss results of the present work in the context of quantum turbulence in dilute Bose gases as described by the Gross-Pitaevskii equation.
We focus, in particular, on non-thermal fixed points identified by means of a strong-wave-turbulence analysis on the basis of two-particle irreducible (2PI) dynamic field equations \cite{Berges:2008wm,Berges:2008sr,Scheppach:2009wu} as well as semi-classical field simulations \cite{Nowak:2010tm,Nowak:2011sk,Schole:2012kt,Schmidt:2012kw,Nowak:2013juc}.

We begin with a brief summary of the results of Ref.~\cite{Scheppach:2009wu}. 
Stationary scaling solutions for the statistical and spectral two-point correlators,
\begin{alignat}{3}
& F\left(s^z \omega,s\mathbf{p}\right) & & = s^{-2-\kappa} {F\left(\omega,\mathbf{p}\right),} \nonumber \\
& \rho\left(s^z \omega,s\mathbf{p}\right) & & = s^{-2+\eta} {\rho\left(\omega,\mathbf{p}\right),}
\label{eq:scaling_bose}
\end{alignat}
respectively, were predicted by means of a non-perturbative wave-turbulence analysis of the 2PI dynamic equations for these correlation functions.

These solutions constitute non-thermal fixed points of the far-from-equilibrium dynamics \cite{Berges:2008wm,Berges:2008sr,Scheppach:2009wu}.
Here, $F$ and $\rho$ are defined in terms of the time-ordered Greens function $G(x-y)=\langle\mathcal{T}\psi^{\dagger}(x)\psi(y)\rangle$ of the on the average translationally invariant Bose gas as $G(x)=F(x)-i\,\mathrm{sgn}(x_{0})\rho(x)/2$, where $x=(t,\mathbf{x})$.
The critical behaviour is characterised by the exponents $\kappa$ and $\eta$, as well as the dynamical exponent $z$.
$\eta$ is an anomalous critical exponent which determines the deviation of the spectral scaling from the free behaviour.
In Ref.~\cite{Scheppach:2009wu} two possible solutions were found, corresponding to different strong-wave-turbulence cascades, with scaling exponents
\begin{alignat}{2}
& \kappa_{\mathrm{P}} = d+2z-\eta_{\mathrm{P}}, 
\nonumber \\
& \kappa_{\mathrm{Q}} = d+z-\eta_{\mathrm{Q}},
\label{eq:NTFPkappa}
\end{alignat}
between $\kappa$, $\eta$, $z$, and $d$.
$(\kappa_{\mathrm{P}},\eta_{\mathrm{P}})$ correspond to an energy cascade while $(\kappa_{\mathrm{Q}},\eta_{\mathrm{Q}})$ reflect a quasi-particle cascade in the wave turbulent system.
Both represent non-thermal fixed points of the non-equilibrium Bose gas \cite{Berges:2008wm,Berges:2008sr,Scheppach:2009wu}.
The scaling of the statistical correlation function $F$ implies  scaling of the single-particle momentum distribution $n(p)=\int d\omega F(\omega,p)$, cf.~\cite{Scheppach:2009wu}:
\begin{align}
n(p)\sim p^{-\zeta},\quad\mbox{with}\quad \zeta=\kappa-z+2.
\label{eq:nscaling2PI}
\end{align}
The scaling of the single-particle kinetic energy, see \Eq{kin_energy}, implies $\zeta=\xi+2 = 2\eta_1-d$.  
Comparing this with \eq{nscaling2PI}, and making use of the constraint $z=2+d-\eta_{1}$, which is a consequence of Galilei invariance, cf.~\Eq{zofeta1}, one obtains $\kappa=\eta_{1}$ and thus, with \Eq{consistency}, the relations
\begin{align}
\begin{array}{l}\kappa_{\mathrm{P}}=\eta_{1},\\ \kappa_{\mathrm{Q}}=\eta_{1},\end{array}
\qquad
\begin{array}{l}\eta_{\mathrm{P}}=d-\eta_{2},\\ \eta_{\mathrm{Q}}=2(d-1-\eta_{2})/3\end{array}
\end{align}
between the exponents of Ref.~\cite{Scheppach:2009wu} and the ones introduced in the present work.
Finally, we make use of \Eqs{zofeta1} and \eq{NTFPkappa} to obtain 
\begin{alignat}{2}
& \kappa_{\mathrm{P}} = d+4/3-\eta_{\mathrm{P}}/3, \nonumber \\
& \kappa_{\mathrm{Q}} = d+1-\eta_{\mathrm{Q}}/2,
\label{eq:NTFPkappawithz}
\end{alignat}
and, with \Eqs{nscaling2PI} and \eq{zofeta1},
\begin{alignat}{2}
& \zeta_{\mathrm{P}} = d+8/3-2\eta_{\mathrm{P}}/3, \nonumber \\
& \zeta_{\mathrm{Q}} = d+2-\eta_{\mathrm{Q}}.
\label{eq:NTFPzetawithz}
\end{alignat}
In turbulence theory, one considers the scaling of the radial kinetic energy distribution $E(p)\sim p^{d-1}\epsilon_{\text{kin}}(p) \sim p^{d-1-\xi}$.
Combining the above results, one finds that {the energy} and particle cascades have radial single-particle kinetic energy distributions 
\begin{alignat}{2}
 E_{\mathrm{P}}(p) &\sim p^{-5/3+2\eta_{\mathrm{P}}/3}, \nonumber \\
 E_{\mathrm{Q}}(p) &\sim p^{-1+\eta_{\mathrm{Q}}},
\label{eq:radialEPQ}
\end{alignat}
respectively.
We find that, for {the energy} cascade, the strong-wave-turbulence scaling \cite{Scheppach:2009wu} of $E_{\mathrm{P}}(p)$ is equivalent to the classical Kolmogorov law \cite{Kolmogorov1941a,*Kolmogorov1941b,*Kolmogorov1941c,Frisch2004a}, with an intermittency correction $2\eta_{\mathrm{P}}/3$.
Kolmogorov-$5/3$ scaling has been reported to be possible in a superfluid both experimentally \cite{Maurer1998,Walmsley2007a,*Walmsley2008a}, and in simulations \cite{Araki2002a,Kobayashi2005a} of the GPE.

Assuming $\eta_{\mathrm{Q}}=0$, the distribution $E_{\mathrm{Q}}(p)$ corresponds, for $d=2,\,3$ to the scaling of the flow field $v\sim r^{-1}$ with the distance $r$ from a vortex core \cite{Nore1997a,*Nore1997b} and, equivalently, of a random distribution of vortices \cite{Nowak:2010tm,Nowak:2011sk}, as we will discuss in more detail in the following.
This inverse cascade plays an important role in the equilibration and condensation process \cite{Svistunov1991a,Semikoz1997a,*Micha2003a} after a strong cooling quench in a Bose gas \cite{Berges:2012us,Nowak:2012gd}.

In Refs.~\cite{Nowak:2010tm,Nowak:2011sk,Schmidt:2012kw,Schole:2012kt,Nowak:2013juc}, the above non-thermal fixed points of the dilute superfluid gas were discussed in the context of topological defect formation and superfluid turbulence.
A key result is that nearly degenerate Bose gases in $d=2,3$ dimensions, quenched parametrically close to the Bose-Einstein condensation (in $d=2$ BKT) transition, can evolve quickly to a quasi-stationary state exhibiting critical scaling \cite{Nowak:2011sk} and slowing-down behaviour \cite{Schole:2012kt}.
The critical scaling exponents $\zeta$ of the single-particle momentum spectra $n(p)\sim p^{-2}\epsilon_{\text{kin}}(p)\sim p^{-\zeta}$ corroborated the predictions $\zeta_\mathrm{Q} = d+2-\eta_\mathrm{Q}$ of the strong-wave-turbulence analysis of Ref.~\cite{Scheppach:2009wu} for a quasi-particle cascade, with a very small value of $\eta_\mathrm{Q}$. 
Within the numerical precision it was found that $\zeta_\mathrm{Q}=4$ in $d=2$ and $\zeta_\mathrm{Q}=5$ in $d=3$ \cite{Nowak:2011sk}.
These exponents turned out to be related to randomly distributed vortices and (large) vortex rings occurring during the approach of the critical state \cite{Nowak:2010tm,Nowak:2011sk}.

Given the relation \eq{radialEPQ}  of the scaling laws with hydrodynamics and topological and geometric properties of the superfluid gas, we call the exponents $-5/3$ and $-1$ canonical while the effects of fluctuations are captured by the anomalous corrections $2\eta_{\mathrm{P}}/3$ and $\eta_{\mathrm{Q}}$, respectively.  

%=======================================================================
\subsection{Acoustic turbulence in a superfluid}
\label{sec:resultsAcoustic}
Let us return to the KPZ dynamics.
In order to make contact with scaling in acoustic turbulence in a superfluid, we insert the average literature values for $\eta_{1}$, as given in Table I. of Ref.~\cite{Kloss2012a}, cf.~also \Fig{eta_range}, which correspond to a forcing potential field delta-correlated in space ($\beta = 2$), into \Eq{kin_energy} and obtain $\epsilon_{\text{kin}}(p) \sim p^{-\xi}$, with
\begin{align}
&\xi=0, & & \mbox{for}\ d = 1, \nonumber \\
&\xi=0.758(30), & & \mbox{for}\ d = 2, \nonumber \\
&\xi=1.600(24), & & \mbox{for}\ d = 3.
\label{eq:epsilon_scaling}
\end{align}
These results can be compared with scaling behaviour observed in acoustic turbulence in ultracold Bose gases, as summarised in the following.

Results related to the quantum turbulence discussed in the previous section were obtained for a $d=1$-dimensional Bose gas in Ref.~\cite{Schmidt:2012kw}. 
There, the relation between critical scaling of the single-particle momentum spectrum and the appearance of solitary wave excitations was pointed out.
It was found that this spectrum, as for a thermal quasi-condensate, has a Lorentzian shape if the solitons are distributed randomly in the system, with the width of the Lorentzian being related to the mean density of solitons.
The latter is in general different from and independent of the thermal coherence length of a gas with the same density and energy.
The kinetic energy spectrum, in the regime of momenta larger than the Lorentzian width, correspondingly shows a momentum scaling $\epsilon_{kin}(p) \sim p^{2} n(p) \sim p^0$.
This, in turn, is in full agreement with the above result quoted in \Eq{epsilon_scaling}, corresponding to a white-noise forcing, i.e., $\beta=2$.
The power law is consistent with that occurring in the single-particle spectrum of a random distribution of grey and black solitons in a one-dimensional Bose gas \cite{Schmidt:2012kw}.

The fixed points found in Ref.~\cite{Kloss2012a}, at which the exponents \eq{epsilon_scaling} apply, describe critical dynamics according to the KPZ equation describing, e.g., the unbounded propagation of an interface moving with coordinates $(\theta,\bf{x})$ in a two-component statistical system. 
On the contrary, the KPZ equation derived for the phase angle $\theta$ of the complex field $\psi$ evolving according to the GPE, see \Sect{B2GPE}, is subject to the additional constraint that the range of angles $0 < \theta \leq 2\pi$ is compact.
This constraint plays an important role if the phase excitations are large enough to allow for (quasi) topological defects.
Hence, one can not expect the predictions \eq{epsilon_scaling} to necessarily match the scalings occurring when defects such as vortices are present.  

To make contact with the scalings \eq{epsilon_scaling}, we note that, while the strong-wave-turbulence prediction $\zeta_\mathrm{Q}=d+2$ is consistent with vortices dominating the infrared behaviour of the single-particle spectrum \cite{Nowak:2011sk}, it does not apply to the $(d=1)$-dimensional case where there are no vortex defects, since $\zeta_\mathrm{Q}=d+2=3$ is by $1$ larger than the exponent $\zeta=2$ appearing in the Lorentzian distribution at large momenta.
However, also in $d=2$ and $d=3$, a scaling $\zeta_{c}\simeq d+1$ appears as a result of kink-like structures and longitudinal, compressible sound excitations.
In Ref.~\cite{Nowak:2011sk}, it was demonstrated that the single-particle spectrum of the compressible component can show power-law behaviour, with an  exponent  $n_{c}(p)\sim p^{-\zeta_{c}}$.
This power-law was ascribed to sound wave turbulence on the background of the vortex gas, in particular to the density depressions remaining for some time in the gas after a vortex and an anti-vortex have mutually annihilated \cite{Berloff2004a}, cf.~Fig.~15 of Ref.~\cite{Nowak:2011sk}.

We compare the predictions  \eq{epsilon_scaling} with the scalings found in \cite{Nowak:2011sk}.
Using $\zeta=2\eta_{1}-d=\xi+2$ one finds
\begin{align}
&\zeta=2, & & \mbox{for}\ d = 1, \nonumber \\
&\zeta=2.758(30), & & \mbox{for}\ d = 2, \nonumber \\
&\zeta=3.600(24), & & \mbox{for}\ d = 3.
\label{eq:n_scaling}
\end{align}
Defining an anomalous exponent $\eta$ by means of the relation
$\zeta=\zeta_{c}-\eta=d+1-\eta$
gives
\begin{align}
&\eta = 0, & & \text{for $d=1$}, \nonumber \\
&\eta = 0.242(30), & & \text{for $d=2$}, \nonumber \\
&\eta = 0.400(14), & & \text{for $d=3$}.
\label{eq:anomalousLit}
\end{align}
%
%=======================================================================
\begin{figure}[t]
\includegraphics[width=0.3\paperwidth]{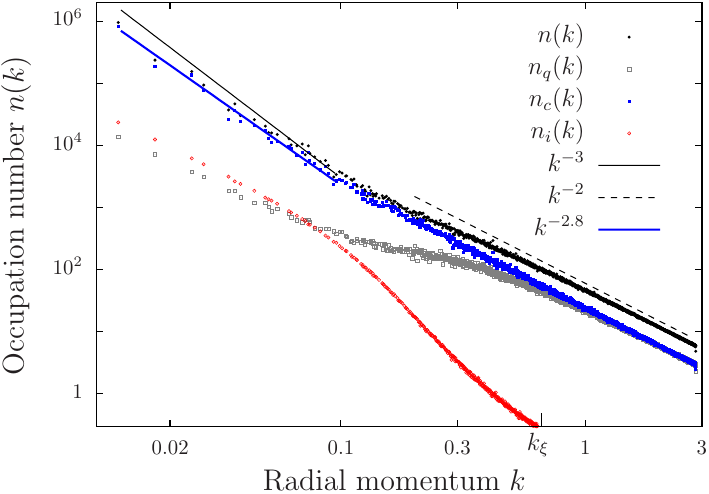}
\includegraphics[width=0.3\paperwidth]{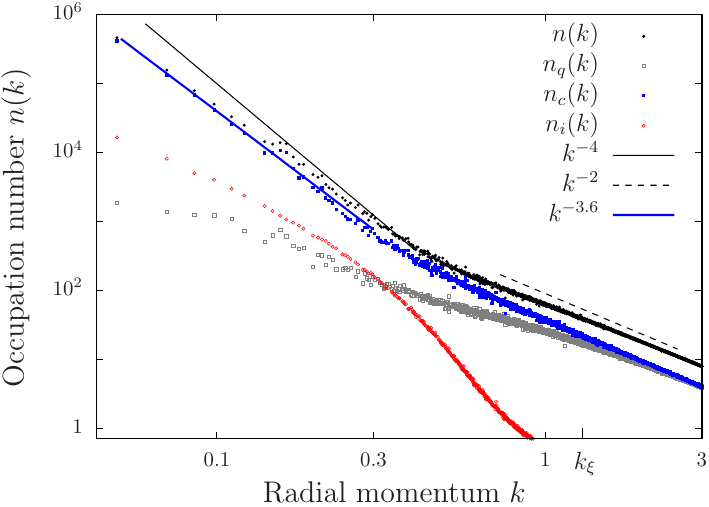}
\caption{(Color online)
Acoustic turbulence in a $d=2$ (upper) and $d=3$ (lower panel) dilute superfluid Bose gas:
Shown are occupation number spectra of the late-stage evolution of a closed system after an initial quench, briefly after the last vortex-antivortex pair ($d=2$) or vortex ring ($d=3$) has disappeared, cf.~Fig.~15 of Ref.~\cite{Nowak:2011sk}.
{The distributions show a snapshot of a time evolution in a closed system.}
The blue solid lines correspond to the predictions obtained in the present work, given in \Eq{n_scaling}, the black solid  lines mark the canonical scaling, $n(p) \sim p^{-d-1}$. 
The latter scaling is expected on geometrical grounds, e.g., results from randomly distributed plane-wave density depression waves or solitons \cite{Nowak:2011sk}.
It is emphasised that the comparatively large deviations ($\sim p^{-2.8}$ for $d=2$ and $\sim p^{-3.6}$ for $d=3$) are found independently and contrast the small deviation from the canonical scalings \eq{radialEPQ} of quantum turbulent spectra in the presence of vortices.
The radial momentum is given in lattice units, $k = [2\sum_{i=1}^d \sin^2(2\pi n_i a/L)]^{1/2}$, with $n_i \in \mathbb{Z}$, $-L/(2a) \leq n_i \leq L/(2a)$,  $a$ being the grid constant and $L$ its side length. 
$k_{\xi} = 2\sin(\pi/(2\xi))$ is the momentum corresponding to the inverse healing length.
}
\label{fig:anomalous_simulations}
\end{figure}
%=======================================================================
For $d=1$, the  scaling \eq{n_scaling} corresponds to that of the Lorentzian of a random soliton gas as discussed above.
Furthermore, within the numerical precision, the power laws seen in Fig.~15 of Ref.~\cite{Nowak:2011sk} are found to be consistent with the values \eq{n_scaling}. 
We reproduce the data in \Fig{anomalous_simulations}, comparing it with the IR scaling exponents \eq{n_scaling} for $d=2, 3$ (blue solid lines). 
The figures show the single-particle occupation number spectrum $n(p) \sim p^{-2}\epsilon_{\text{kin}}(p)$ obtained from a numerical simulation of the GPE.
The particular scalings occur shortly after the decay of the last topological excitations, i.e., the last vortex-antivortex pair for $d=2$ or vortex ring for $d=3$. 
At the time the picture is taken, the compressible excitations dominate and their scaling exponent can be measured. 
The relatively large anomalous predictions of \Eq{anomalousLit}  fit the data very well.

We remark that the grey solitary-wave excitations as well as the density depressions remaining after vortex-anti-vortex annihilation are consistent with the absence of the compactness constraint on $\theta$ in the KPZ equation.
The soliton gas can be dominated by grey solitons \cite{Schmidt:2012kw} which imply only weak density depressions at the position of the phase jump.
The weaker the depression, the smaller the phase kink and the less relevant the compactness of the range of possible $\theta$.
Similarly the density depressions leading to $\zeta_{c}\simeq d+1$ do not require the phase to vary over the full circle.
Hence, we expect in these cases that KPZ predictions for critical exponents apply also to the GPE, as  defects do not play a role.

%=======================================================================
%=======================================================================
\section{Summary}
\label{sec_concl}
In this article, we have investigated the scaling properties of the stochastic Burgers equation within the functional renormalisation group (RG) approach. 
For this purpose, a set of fixed-point equations has been established for the two-point correlators, including, in particular, the velocity cumulants.
They are solved in both, the ultraviolet and infrared asymptotic regimes.
A regime of UV-convergent fixed points was identified.
It was found that direct cascades of energy require a UV regulator as it is set by a dissipation term. 
On the other hand, the IR asymptotic form of the integral equations was found to allow for a continuous set of solutions.
In one spatial dimension, the continuous set of fixed points was shown to coincide with perturbative results for the KPZ equation \cite{Medina1989a}. 

Moreover, we found interesting implications for strong wave and quantum turbulence, as well as acoustic turbulence in a dilute Bose gas.
In particular, we have shown that the canonical Kolmogorov-type $5/3$ exponent can be derived for {the energy} cascade in a quantum system from previously studied strong-wave-turbulence solutions.
On the basis of existing numerical studies we hence conjecture the corresponding anomalous exponents of superfluid turbulence to be close to zero.
Using literature averages for KPZ critical exponents we furthermore obtained first quantitative estimates for the anomalous scaling of acoustic turbulence in a superfluid Bose gas, corresponding to ensembles of grey solitons in one-dimensional condensates and to density depressions and sound wave turbulence in two- and three-dimensional systems. These anomalous exponents are corroborated by independent existing numerical simulations.

%==============================================================================
%
\acknowledgements
We thank E. Altman, J. Berges, S. Bock, L. Ca\-net, I. Chantesana, S. Diehl, S. Erne, M. Karl, T. Kloss, A. Liluashvili, D. Mesterhazy, M. Mitter, M. M\"uller,  M. K. Oberthaler, A. Samberg, C. Wetterich, and N. Wschebor for discussions, and B. Nowak and J. Schole for simulations related to \Fig{anomalous_simulations}.
This work was supported by the Deut\-sche Forschungsgemeinschaft (GA677/7,8), the European Commission (ERC-AdG-290623), the Helmholtz Association (HA216/EMMI), and the University of Heidelberg (LGFG and Center for Quantum Dynamics).

%=======================================================================
%=======================================================================
\appendix

%=======================================================================
%=======================================================================
\section{Superfluid hydrodynamics}
\label{app:superfluid_hydro}
In the following we give details on the relation  between the driven dissipative Gross-Pitaevskii equation \eq{GPE} and the stochastic Burgers equation \eq{Burgers}.
{These are related through} the hydrodynamic decomposition of the complex field, {$\psi = \sqrt{n}\exp\{\I \theta\}$},
\begin{alignat}{3}
& \partial_t \theta + \frac{1}{2m}\left( \boldsymbol{\nabla}\theta \right)^2 - \nu \nabla^2 \theta &&= U, \nonumber \\
& \partial_t n + \frac{1}{m} \boldsymbol{\nabla} \cdot \left( n \boldsymbol{\nabla}\theta \right) &&= S.
\end{alignat}
This is formally similar to the equations that arise from the conservative GPE, with the addition that the continuity equation is in-homogeneous and that the KPZ equation has a non-zero dissipative term,
\begin{alignat}{3}
& U &&= \frac{1}{4 m \sqrt{n}} \boldsymbol{\nabla} \cdot \left(\frac{\boldsymbol{\nabla}n}{\sqrt{n}}\right) + \frac{\nu}{n}\boldsymbol{\nabla}n \cdot \boldsymbol{\nabla}\theta \nonumber \\
&&& \quad +\mu_1 - g_1 n - \frac{\text{Re}(\zeta{\rm e}^{-i\theta})}{\sqrt{n}}, \nonumber \\
& S && = \nu \sqrt{n} \boldsymbol{\nabla} \cdot \left(\frac{\boldsymbol{\nabla}n}{\sqrt{n}}\right) - 2 \nu n \left(\boldsymbol{\nabla}\theta\right)^2 \nonumber \\
&&& \quad - 2\mu_2 n - 2 g_2 n^2 + 2 \sqrt{n} \text{Im}(\zeta{\rm e}^{-i\theta}).
\end{alignat}
These equations are coupled non-linear Langevin equations.
If the fluctuations of the field amplitude are subdominant the former can be decoupled by assuming that $U$ plays the role of the potential of the stochastic forcing {${\bf f} = m^{-1}\boldsymbol{\nabla}U$}, with noise correlator
\begin{align}
\langle U(\omega,\mathbf{p}) U(\omega',\mathbf{p}') \rangle 
= {\delta(\omega+\omega')\, \delta\left(\mathbf{p}+\mathbf{p'}\right) \, u(\omega,\mathbf{p}).}
\label{eq:forcingKPZ}
\end{align}
This describes particles being injected and removed as amplitude fluctuations, such that the system reaches a state where they can be described by a (not necessarily thermal) distribution and feed energy to the phase fluctuations. Note that, contrarily to the correlations of $\zeta$, we do not require $U$ to be delta correlated in space. Burgers equation is obtained by setting {${\bf v} = m^{-1}\boldsymbol{\nabla}\theta$}.

The kinetic energy spectrum is defined in terms of the two-point correlation function of $\psi$. 
It can be decomposed into three parts,
\begin{align}
 E_{\text{kin}} &\equiv -\frac{1}{2m} \int_{\bf x} \langle \psi^{\dagger} \nabla^2\psi \rangle 
\nonumber\\
&= \frac{\rho}{2m} \int_{\bf x} \langle \left(\boldsymbol{\nabla}\theta\right)^2 \rangle 
+  \frac{1}{2m} \int_{\bf x} \langle \frac{\left(\boldsymbol{\nabla}n\right)^2}{4 n} \rangle \nonumber \\
 & \quad {+ \frac{1}{2m}\int_{\bf x} \, {\langle \delta n \left(\boldsymbol{\nabla}\theta\right)^2 \rangle}} \nonumber \\
&= E_{\text{phase}} + E_{\text{amplitude}} + E_{\text{exchange}}.
\end{align}
The amplitude of $\psi$ is separated into $n = \langle n \rangle + \delta n \coloneqq \rho  + \delta n$. 
At sufficiently low energies, the average value of the amplitude is much larger than its fluctuations and the major contribution to the kinetic energy is $E_{\text{phase}}$. 
Then,
\begin{align}
 E_{\text{kin}} \cong \frac{\rho}{2m} \int_{\bf x} \, \langle \left(\boldsymbol{\nabla}\theta\right)^2\rangle \coloneqq \mathcal{V} \int_{\bf p} \epsilon_{\text{kin}}({\bf p}),
\end{align}
where $\mathcal{V}$ is the volume of the system. 
Hence,
\begin{align}
 \epsilon_{\text{kin}}({\bf p})&= \frac{m \rho}{2} \int_{\omega} \, \langle {\bf v}(\omega,{\bf p}) \cdot {\bf v}(-\omega,-{\bf p})\rangle.
\end{align}
%
%=======================================================================
%=======================================================================

\section{Effective action and observables}
\label{sec:defgamma}
In the following we give details of the implementation of the coarse graining by means of the cutoff function $R_{k}$.

The Schwinger functional at scale $k$ is defined as
\begin{alignat}{3}
& \text{e}^{ W_k[\mathbf{J}]} & &= \int \underset{p>k}{\Pi} \, \text{d}\mathbf{v}(\omega,\mathbf{p}) \, \text{e}^{-S[\mathbf{v}]+\int_{t,\mathbf{x}} \, \mathbf{J}(t,\mathbf{x}) \cdot \mathbf{v}(t,\mathbf{x})} \nonumber \\
&&& = \int D\mathbf{v} \, \text{e}^{-S[\mathbf{v}]- \Delta S_k[\mathbf{v}] +\int_{t,\mathbf{x}} \, \mathbf{J}(t,\mathbf{x}) \cdot \mathbf{v}(t,\mathbf{x})}, \\
&\Delta S_k[\mathbf{v}] & &= \frac{1}{2}\int_{\omega,\mathbf{p}} \mathbf{v}(\omega,\mathbf{p}) \cdot \mathbf{v}(-\omega,-\mathbf{p}) \, R_k(p),
\end{alignat}
with $R_k(p)>0$, $R_k(p\gg k) = 0$ and $R_k(p\ll k) = \infty$. 
$\Gamma_{k\to 0}[\mathbf{v}] \equiv \Gamma[\mathbf{v}]$ is the Legendre transform of the Schwinger functional, $W[\mathbf{J}]$ which generates all the correlation functions of the velocity field. 
This information is equivalently contained in $\Gamma[\mathbf{v}]$ which allows to derive correlation functions. 
For this one determines the average field $\bar{\mathbf{v}}=\langle \mathbf{v}(t,\mathbf{x}))\rangle$ by solving the equation of motion $\delta \Gamma /\delta \mathbf{v}[\bar{\mathbf{v}}]=0$. In our case, we get $\bar{\mathbf{v}}=0$. 
See, e.g., Ref.~\cite{Zinnjustin2002a} for {the} more general cases. 
The two-point correlation function is the inverse of the second derivative of $\Gamma[\mathbf{v}]$ evaluated at $\bar{\mathbf{v}}$,
\begin{align}
\langle v_i(t,\mathbf{x}) v_j(t',\mathbf{x}') \rangle = \left( \left. \frac{\delta^2 \Gamma}{\delta v_i(t,\mathbf{x}) \delta v_j(t',\mathbf{x}') }\right|_{\bar{\mathbf{v}}} \right)^{-1}.
\end{align}
The inverse is defined through
\begin{align}
\int_{\tau,\mathbf{z}} \,& \langle v_i(t,\mathbf{x}) v_k(\tau,\mathbf{z}) \rangle \left. \frac{\delta^2 \Gamma}{\delta v_k(\tau,\mathbf{z}) \delta v_j(t',\mathbf{x}') }\right|_{\bar{\mathbf{v}}} \nonumber \\
&  \qquad  = \delta_{i,j} \, \delta(t-t') \, \delta(\mathbf{x}-\mathbf{x}').
\end{align}
{Higher order moments are computed from the higher order derivatives of $\Gamma[\mathbf{v}]$ multiplied by external legs.}

%=======================================================================
%=======================================================================
\section{RG flow equations}
\label{ap_I}
In this appendix we give explicit expressions for the integral terms in the RG flow equations \eq{flow1}. 
We use the sharp cutoff $R_{k;ij}(p) = \delta_{ij} \, k^d {z_1}\tilde R_{k}(p)$, with
\begin{align}
\tilde R_{k}(p)  
\left\{ \begin{array}{ll}
=0 & \text{if } p \ge k, \\
\to\infty & \text{if } p < k. 
\end{array}\right.
\end{align}
As a result we can use the identity
\begin{align}
& \frac{1}{\Gamma_k^{(2)}+R_k} k\frac{\partial R_k}{\partial k} \frac{1}{\Gamma_k^{(2)}+R_k} = 2k^2 \delta(p^2-k^2) \frac{1}{\Gamma_k^{(2)}}
\label{eq:identity}
\end{align}
to evaluate the flow integrals.
To proceed, we read off, from the ansatz \eq{trunc}, the two-point function
\begin{align}
 & \frac{\delta^2 \Gamma_k[v = 0]}{\delta v_i(\omega',\vec{p}') \delta v_j(\omega,\vec{p})} 
 = \frac{\Gamma_{k,ij}^{(2)}(\omega,\mathbf{p})}{(2\pi)^{d+1}} \delta(\omega+\omega') \delta\left(\mathbf{p}+\mathbf{p}'\right),
\label{eq:Gamma2}
\end{align}
with $\Gamma_{k,ij}^{(2)}(\omega,\mathbf{p})$ defined in \Eq{Gprop}, the 3-vertex
\begin{align}
&  \frac{\delta^3 \Gamma_k[v=0]}
{\delta v_i(\omega',\mathbf{p}') \delta v_j(\omega,\mathbf{p}) \delta v_l(\omega'',\vec{q})} 
= {(2\pi)^{-2(d+1)}}
\nonumber \\
& \ \times\ {\delta(\omega+\omega'+\omega'')\, \delta\left(\mathbf{p}+\mathbf{p}'+\mathbf{q}\right)} \,\Gamma_{k;ijl}^{(3)}(\omega',\vec{p}';\omega,\vec{p}),
\end{align}
with
\begin{align}
& \Gamma_{k;ijl}^{(3)}(\omega',\vec{p}';\omega,\vec{p})  
 = -F^{-1}_k(\left|\vec{p}+\vec{p}'\right|) 
\nonumber\\
& \qquad\times\ 
\left(\omega +\omega' +i\nu_k(\left|\vec{p}+\vec{p}'\right|) \left|\vec{p}+\vec{p}'\right|^2\right) 
\left[\delta_{il}p'_j + \delta_{jl}p_i \right] 
\nonumber\\
&\quad -\ F^{-1}_k(p') \left(-\omega' +i\nu_k(p') \, p'^2\right) 
\left[\delta_{ij}p_l-\delta_{il}(p'_j+p_j)\right] 
\nonumber \\
&\quad -\ F^{-1}_k(p) \left( -\omega +i\nu_k(p) \, p^2 \right) 
\left[\delta_{ij}p'_l -\delta_{jl}(p'_i+p_i)\right],  
\label{eq:Gamma3}
\end{align}
and the 4-vertex
\begin{align}
 &\frac{\delta^4 \Gamma_k[v=0]}
 {\delta v_i(\omega,\mathbf{p}) \delta v_j(\omega',\mathbf{p'}) 
  \delta v_l(\omega'',\mathbf{q}) \delta v_m(\omega''',\mathbf{q'})}
 ={(2\pi)^{-3(d+1)}} 
 \nonumber \\
& \qquad \times\
{\delta(\omega+\omega'+\omega''+\omega''') \delta\left(\mathbf{p}+\mathbf{p}'+\mathbf{q}+\mathbf{q}'\right)} \nonumber \\
& \qquad \times \Gamma_{k,ijlm}^{(4)}(\mathbf{p},\mathbf{p}',\mathbf{q}),
\end{align}
with
\begin{align}
 &\Gamma_{k,ijlm}^{(4)}(\mathbf{p},\mathbf{p}',\mathbf{q})  
 \nonumber \\
&\ =  F_k^{-1}(\left|\mathbf{p}+\mathbf{p}'\right|) 
\big[\left(\delta_{mi} p_j + \delta_{mj} p'_i \right) \left(p_l + p'_l + q_l\right)
\nonumber \\
&\qquad\qquad\qquad\quad
-\left(\delta_{li} p_j + \delta_{lj} p'_i \right) q_m\big] 
\nonumber \\
& \quad + F_k^{-1}(\left|\mathbf{p}+\mathbf{q}\right|)
\big[\left(\delta_{im} p_l+\delta_{lm}q_i\right) \left(p_j + p'_j + q_j\right) 
\nonumber \\
& \qquad\qquad\qquad\quad 
-\  \left(\delta_{ij}p_l  + \delta_{jl} q_i\right)p'_m \big] 
\nonumber \\
& \quad + F_k^{-1}(\left|\mathbf{p}'+\mathbf{q}\right|)
\big[\left(\delta_{jm}p'_l +\delta_{lm}q_j\right)\left( p_i + p'_i + q_i \right)
\nonumber \\
& \qquad\qquad\qquad\quad 
-\ \left(\delta_{ij} p'_l  +\delta_{il} q_j\right)p_m \big].
\label{eq:Gamma4}
\end{align}
Using these, the flow integral \eq{flow_3} can be written as
\begin{align}
& I_{k}^{(2)}[0](\omega,\mathbf{p}) 
 = \frac{k^{2}}{d} \int_{\omega',\mathbf{p'}} \delta(p'^2-k^2) {\Gamma_{k;ni}^{(2)}}^{-1}(\omega',\vec{p}')
 \nonumber\\
 &\times \Big[ 2 \, {\Gamma_{k,jm}^{(2)}}^{-1}(\omega'-\omega,\vec{p}'-\vec{p})
\theta ( (\vec{p}'-\vec{p})^2 -k^2) \nonumber\\
&\qquad\times\
\,{\Gamma_{k;ijl}^{(3)}(-\omega',-\vec{p}';\omega'-\omega,\mathbf{p}'-\vec{p})} \nonumber \\
&\qquad\times\ \Gamma_{k;mnl}^{(3)}(\omega-\omega',\vec{p}-\vec{p}';\omega',\vec{p}') \, 
\nonumber\\
&\quad
  -\ {\Gamma_{k;nijj}^{(4)}(\mathbf{p},-\mathbf{p},\mathbf{p}')}  
\Big].&
\label{eq:I}
\end{align}
The theta function arises because of the sharp cutoff.
Since the cutoff diverges for $p<k$, the propagator $G_{k}$ vanishes in this regime, cf. \Eq{Gk}. 
This is irrelevant in the diagram depending on the 4-point vertex, cf.~\Eq{flow_3},  where the single appearing propagator carries a $\partial_{k}R_{k}$ insertion and is thus evaluated at $p^{2}=k^{2}$.
Because of rotational symmetry, $\delta^2 I_{k}/\delta v_{i}^2[0](\omega,\mathbf{p})$ does not depend on $i$.
Hence, taking the trace of $\delta^2 I_{k}/\delta v_{i} \delta v_{j}[0](\omega,\mathbf{p})$ and dividing it by $d$ makes the  integrals rotationally invariant.

Because of \eq{Gprop}, the integrand is a rational function of $\omega'$ such that the $\omega$ integration can be done.
Introducing the short-hand notation $\nuk{p}{}=\nu_{k}(p)p^{2}$, etc., we obtain 
\begin{align}
& I_{k}^{(2)}[0](\omega,\mathbf{p}) 
= -\frac{k^2}{d} \int_{\mathbf{q},\mathbf{r}} 
{(2\pi)^{d}\delta(r^2-k^2) \,  \delta(\mathbf{p}-\mathbf{q}-\mathbf{r})}
\nonumber\\
&\times 
\big\{F^{-1}_k(q) F^{-1}_k(r) \nukq \nukr [\omega^2 + (\nukq + \nukr)^2]\big\}^{-1}
\nonumber \\
& \times \Big[ 
 F^{-1}_k(q)^2 \nukq \left[\omega^2 + (\nukq + \nukr)^2\right]
\left[d(p^2+r^2) - 2 \mathbf{p} \cdot \mathbf{r}\right]
\nonumber \\
& \quad - \theta \left( q^2-k^2\right) 
\nonumber\\
& \quad\times\Big( 
F^{-1}_k(q)^2 \nukq \left[\omega^2 + (\nukq + \nukr)^2\right]
\left[d(p^2+r^2)-2\mathbf{p}\cdot\mathbf{r}\right] 
\nonumber \\
& \quad\ \ + F^{-1}_k(r)^2 \nukr \left[\omega^2 + (\nukq + \nukr)^2\right] 
\left[d(p^2+q^2)-2\mathbf{p} \cdot \mathbf{q}\right] 
\nonumber \\
& \quad\ \ + F^{-1}_k(p)^{2}  \left(\nukq + \nukr\right) \left[\omega^2 + \nukps\right]
\left[d (q^2 + r^2) + 2\mathbf{q} \cdot \mathbf{r} \right]
\nonumber \\
& \quad\ \ + F^{-1}_k(p)F^{-1}_k(q) \nukq \left[ \omega^2 -  \nukp\left(\nukq + \nukr\right) \right] 
2d\,\mathbf{p}\cdot \mathbf{q}
\nonumber \\
& \quad\ \ + F^{-1}_k(p)F^{-1}_k(r) \nukr \left[ \omega^2 -  \nukp\left(\nukq + \nukr\right) \right] 
2d\,\mathbf{p}\cdot \mathbf{r}
\Big)\Big].
\end{align}
Note that the terms $\propto d$ and independent of $d$ originate from contractions of the types $\delta^{ij}\delta^{ji}p^k  q^k$ and $\delta^{ij}  \delta^{jl}  p^i  q^l$, respectively.
We can integrate radially, $\int_{\vec{r}} = \int_0^\infty r^{d-1} dr \int_{\Omega}$, which, for $d=1$, reduces to $\int_{\Omega} {f(p)} = [f(p)+f(-p)]/(2\pi)$.
{The delta distributions allow to set $\mathbf{r} = k \mathbf{e}_{\mathbf{r}}$ and $\mathbf{q} = \mathbf{p}-k\mathbf{e}_{\mathbf{r}}$, with $|\hat{\mathbf{r}}|=|\mathbf{e}_{\mathbf{r}}|=1$.} 
Expansion in powers of $\omega^2$ gives
\begin{align}
& I_{k}^{(2)}[0](0,\mathbf{p}) 
= -\frac{k^d}{2d} \int_{\Omega} \left[F^{-1}_k(k) F^{-1}_k(q) \nukk \nukq (\nukk+\nukq)\right]^{-1} 
\nonumber \\
& \times \Big[ F^{-1}_k(q)^2 \nukq (\nukq+\nukk) 
   \left[ d(p^2+k^2)-2 k \mathbf{e}_{\mathbf{r}} \cdot \mathbf{p}\right]  
   \nonumber \\
& \quad -\ \theta \left(q^2-k^2 \right)
\nonumber \\
%\end{align}
%
%
%\begin{align}
& \quad\times\Big( F^{-1}_k(q)^{2} \nukq (\nukk+\nukq) \left[d(p^2+k^2)-2k\, \mathbf{p}\cdot\mathbf{e}_{\mathbf{r}}\right]
\nonumber\\
&  \quad\ \ + F^{-1}_k(k)^{2} \nukk(\nukk+\nukq) \left[d(p^2+q^2)-2\mathbf{p}\cdot \mathbf{q}\right]
\nonumber\\
&  \quad\ \ + F^{-1}_k(p)^2  \left.\nukp\right.^2 \left[d(k^2+q^2)+2k\, \mathbf{q}\cdot\mathbf{e}_{\mathbf{r}}\right]
\nonumber\\
&  \quad\ \ - F^{-1}_k(p)F^{-1}_k(q) \nukp\nukq  2d \, \mathbf{p} \cdot \mathbf{q} 
\nonumber \\
&  \quad\ \ - F^{-1}_k(p)F^{-1}_k(k) \nukp\nukk 2d \, k\,\mathbf{p}\cdot\mathbf{e}_{\mathbf{r}} \Big) \Big]\, ,
\label{eq:explicit_flow_1}
\end{align}
and
\begin{align}
& \left. {\partial_{\omega^2}} I_{k}^{(2)}[0]\right|_{(0,\mathbf{p})} 
= \frac{k^d}{2d} \int_{\Omega} \theta\left(q^2-k^2\right)F^{-1}_k(p)
\nonumber \\
&\quad\times\  \left[F^{-1}_k(k) F^{-1}_k(q) \nukk \nukq (\nukk + \nukq)^{3} \right]^{-1}
\nonumber\\
& \times \Big( 
F^{-1}_k(p) \, \left[(\nukk+\nukq)^{2} - \left. \nukp \right.^2  \right] \left[d(q^2+k^2)+2 k\,\mathbf{q}\cdot\mathbf{e}_{\mathbf{r}}  \right]  
\nonumber \\
& \quad + 
F^{-1}_k(q) \, \nukq \left[\nukk+\nukq + \nukp  \right]  2d \, \mathbf{p} \cdot \mathbf{q} 
\nonumber \\
& \quad + 
F^{-1}_k(k) \, \nukk \left[\nukk+\nukq + \nukp \right]  2d \, k\,\mathbf{p}\cdot \mathbf{e}_{\mathbf{r}}   
\Big).
\label{eq:explicit_flow_2}
\end{align}
%

%=======================================================================
%=======================================================================
\section{Scaling analysis of flow integrals}
\label{app:ScalingAnalysisFI}
%
%=======================================================================
\subsection{Classical limit ($p\ll k$)}
To analyze the scaling of the flow integrals in the limit $p\ll k$ we go over to  rescaled variables  $\hat{p}=p/k$, $\hat{\mathbf{q}} = \mathbf{q}/k = \hat{\mathbf{p}}-\mathbf{e}_{\mathbf{r}}$, cf.~\Eq{dim_variables}, and insert the parametrisation \eq{nuOm_2_dZ} into Eqs.~\eq{explicit_flow_1} and \eq{explicit_flow_2}.
This allows to define dimensionless flow integrals
\begin{align}
& \hat{I}^{(2)}_1(\hat{p}) 
\equiv k^{-d} \sqrt{\frac{z_1}{z_2}}  I_{k}^{(2)}[0](0,k \, \hat{\mathbf{p}}), \nonumber \\
& \hat{I}^{(2)}_2(\hat{p}) 
\equiv k^{-d+4} \left(\frac{z_1}{z_2}\right)^{3/2} \left. {\partial_{\omega^2}} I_{k}^{(2)}[0]\right|_{(0,k \, \hat{\mathbf{p}})},
\label{eq:I12hat}
\end{align}
where the indices $1,2$ now denote the 0th and 1st-order derivatives with respect to $\omega^{2}$.
We split the integrals $\hat{I}_1^{(2)}$ and $\hat{I}_2^{(2)}$ into three and four parts, respectively, according to their leading scaling behaviour in the limit $\hat{p} \ll 1$ of non-coarse-grained IR momenta.
Introducing the short-hand notation 
\begin{align}
S_i(\hat p) = \sqrt{1+\delta Z_i(\hat p)},
\end{align}
the rescaled flow integrals take the form
\begin{align}
\hat{I}^{(2)}_1(\hat{p}) 
&= \hat{p} \, F_{1,1}(\hat{p}) 
\nonumber\\
& + \hat{p}^{(\eta_1+\eta_2)/2+2} S_1(\hat{p}) \, S_2(\hat{p}) \, F_{1,2}(\hat{p})
\nonumber\\
& + \hat{p}^{\eta_1+\eta_2+2\,\delta_{d1}} [S_1(\hat{p}) \, S_2(\hat{p})]^2 F_{1,3}(\hat{p}), 
\label{eq:rescaled_flow_1}
\\
 \hat{I}^{(2)}_2(\hat{p}) 
 &= \hat{p}^{(\eta_1+\eta_2)/2+2} S_1(\hat{p}) \, S_2(\hat{p}) \, F_{2,1}(\hat{p}) 
 \nonumber \\
 &+ \hat{p}^{\eta_2+2} S_2(\hat{p})^2 F_{2,2}(\hat{p}) 
 \nonumber \\
 &+ \hat{p}^{\eta_1+\eta_2 + 2\,\delta_{d1}} [S_1(\hat{p}) \, S_2(\hat{p})]^2 F_{2,3}(\hat{p})
 \nonumber \\
 &+ \hat{p}^{2\eta_2+2 \, \delta_{d1}} S_2(\hat{p})^4 F_{2,4}(\hat{p}).
\label{eq:rescaled_flow_2}
\end{align}
Using, furthermore, the abbreviation $(\nukk+\nukq)^{-1} = k^{-2} (z_2/z_1)^{1/2} \T$, the functions $F_{i,j}(\hat{p})$ are given by
\begin{align}
 F_{1,1}(\hat{p}) 
 =& \frac{1}{2d\,\hat{p}} \int_{\Omega} 
\nonumber \\
& \times \Big\{\theta \left(\hat{q}^2-1\right) 
\hat{q}^{-(\eta_1+\eta_2)/2} 
S_2(1)^2 [S_1(\hat{q})S_2(\hat{q})]^{-1} \nonumber \\
& \quad \quad \times\left[d (\hat{p}^2 + \hat{q}^2)-2\hat{\mathbf{p}}\cdot\hat{\mathbf{q}}\right] \nonumber \\
& \quad -\theta \left(1-\hat{q}^2\right) \hat{q}^{\eta_2} S_2(\hat{q})^2 [S_1(1)S_2(1)]^{-1} \nonumber \\
& \quad \quad \times \left[d(\hat{p}^2+1)-2 \, \hat{\mathbf{p}} \cdot\mathbf{e}_{\mathbf{r}} \right]  \Big\},
\label{eq:F11}
\end{align}
\begin{align}
 F_{1,2}(\hat{p}) 
=& -\hat{p}^{-2} \int_{\Omega} \theta \left(\hat{q}^2-1\right) \T \big\{[S_1(1)S_2(1)]^{-1} 
\hat{\mathbf{p}} \cdot \hat{\mathbf{q}}  \nonumber \\
&  + \hat{q}^{-(\eta_1+\eta_2)/2} [S_1(\hat{q})S_2(\hat{q})]^{-1}
\hat{\mathbf{p}} \cdot\mathbf{e}_{\mathbf{r}} \big\},
\label{eq:F12}
\end{align}
\begin{align}
F_{1,3}(\hat{p}) 
=&\ \frac{\hat{p}^{-2\delta_{d1}}}{2d} \int_{\Omega} \theta \left(\hat{q}^2-1 \right) \T
\left[d(\hat{q}^2+1) + 2 \hat{\mathbf{q}} \cdot\mathbf{e}_{\mathbf{r}}\right]
\nonumber \\
& \times [S_1(\hat{q})S_2(\hat{q})S_1(1)S_2(1)]^{-1} 
\hat{q}^{-(\eta_1+\eta_2)/2}, 
\label{eq:F13}
\end{align}
\begin{align}
F_{2,1}(\hat{p}) 
=&\  \hat{p}^{-2} \int_{\Omega} \theta \left(\hat{q}^2-1\right)\left.\T \right.^3 
\big\{[S_1(1)S_2(1)]^{-1}\hat{\mathbf{p}} \cdot \hat{\mathbf{q}}  
\nonumber \\
& +\ \hat{q}^{-(\eta_1+\eta_2)/2} [S_1(\hat{q})S_2(\hat{q})]^{-1}
\hat{\mathbf{p}} \cdot\mathbf{e}_{\mathbf{r}}\big\},
\label{eq:F21}
\end{align}
\begin{align}
 F_{2,2}(\hat{p}) 
 =&\ \hat{p}^{-2} \int_{\Omega} \theta \left( \hat{q}^2-1\right) \left.\T \right.^2 
 \big\{[S_1(1)S_2(1)]^{-1}  \hat{\mathbf{p}} \cdot \hat{\mathbf{q}}  
 \nonumber \\
& +\ \hat{q}^{-(\eta_1+\eta_2)/2} [S_1(\hat{q})S_2(\hat{q})]^{-1} 
\hat{\mathbf{p}} \cdot\mathbf{e}_{\mathbf{r}} \big\},
\label{eq:F22}
\end{align}
\begin{align}
F_{2,3}&(\hat{p}) 
= - \frac{\hat{p}^{-2\delta_{d1}}}{2d} \int_{\Omega} \theta \left( \hat{q}^2-1 \right) \left.\T \right.^3 
\hat{q}^{-(\eta_1+\eta_2)/2} 
\nonumber \\ 
& \times (S_1(\hat{q})S_2(\hat{q}) S_1(1)S_2(1))^{-1} 
\left[d(\hat{q}^2+1) + 2 \hat{\mathbf{q}} \cdot\mathbf{e}_{\mathbf{r}}\right],
\label{eq:F23}
\\
F_{2,4}&(\hat{p}) = F_{1,3}(\hat{p}).
\label{eq:F24}
\end{align}
Recall that, in the above expressions, $\hat{q}= |\hat{\mathbf{p}}-\mathbf{e}_{\mathbf{r}}|$.
Hence, in integer $d>1$, $\hat{q}^{2}=\hat{p}^{2}+1-2\hat{p}\cos\theta$, and $\hat{\mathbf{p}} \cdot\mathbf{e}_{\mathbf{r}}=\hat{p}\cos\theta$, $\hat{\mathbf{q}} \cdot\mathbf{e}_{\mathbf{r}}=\hat{p}\cos\theta-1$, $\hat{\mathbf{p}} \cdot\hat{\mathbf{q}}=\hat{p}^{2}-\hat{p}\cos\theta$, where $\theta$ is the angle between $\hat{\mathbf{p}}$ and $\hat{\mathbf{r}}=\mathbf{e}_{\mathbf{r}}$. 
In $d=1$, one has $\hat{q}^{2}=(\hat{p}-1)^{2}$, etc.
{The different terms in \Eqs{rescaled_flow_1}, \eq{rescaled_flow_2} are defined in such a way that the functions $F_{i,j}(\hat{p})$ are analytic and non-vanishing at $\hat{p}=0$, i.e., can be expanded as $F_{i,j}(\hat{p}\to 0) = F_{i,j}(0) + F_{i,j}'(0) \, \hat{p} + \text{O}(\hat{p}^2)$. 
{Note that, for $d=1$, an additional factor $\hat{p}^2$ has to be extracted from $F_{1,3}(\hat{p})$, $F_{2,4}(\hat{p})$, and $F_{2,3}(\hat{p})$ because in this case $d(\hat{q}^2+1) + 2 \mathbf{e}_{\mathbf{r}} \cdot \hat{\mathbf{q}} = \hat{p}^2$. 
We have taken this into account by inserting Kronecker deltas $\delta_{d1}$.}
Then, in the limit $\hat{p}\to 0$, where $\delta Z_i(\hat{p})$ behaves as in \Eq{small_p}, i.e., where
\begin{align}
S_i(\hat{p}\rightarrow 0) = \sqrt{a_i}\, \hat{p}^{(-\eta_i+\alpha_i)/2}\left[1 + \mathcal{F}_{i}\left(\hat{p}\right) \right]^{1/2},
\label{eq:Ssmall_p}
\end{align}
one can use that $\mathcal{F}_{i}\left(\hat{p}\to0\right)$ contains sub-dominant terms, involving possibly logarithms. 
Neglecting these terms, 
\begin{align}
& S_i(\hat{p}\rightarrow 0) \cong \sqrt{a_i}\, \hat{p}^{(-\eta_i+\alpha_i)/2} \nonumber \\
& F_{i,j}(\hat{p}\to 0) \cong F_{i,j}(0),
\end{align}
in Eqs.~\eq{rescaled_flow_1} and \eq{rescaled_flow_2} gives the asymptotic relations}
\begin{align}
\hat{I}^{(2)}_1&(\hat{p}\ll 1) 
\cong \hat{p} \, F_{1,1}(0) + \sqrt{a_1 \, a_2} \, \hat{p}^{(\alpha_1+\alpha_2+4)/2} \, F_{1,2}(0) \nonumber \\
& +\ a_1 a_2 \, \hat{p}^{\alpha_1+\alpha_2+2\,\delta_{d1}} F_{1,3}(0)
\label{eq:IRscalingI1}
\\
\hat{I}^{(2)}_2&(\hat{p}\ll 0)
\cong \sqrt{a_1 \, a_2} \, \hat{p}^{(\alpha_1+\alpha_2+4)/2} \, F_{2,1}(0) \nonumber \\
& +\ a_2 \, \hat{p}^{\alpha_2+2} F_{2,2}(0) 
   +  a_1  a_2 \, \hat{p}^{\alpha_1+\alpha_2 + 2\,\delta_{d1}} \, F_{2,3}(0) \nonumber \\
& +\ a_2^2 \, \hat{p}^{2\alpha_2+2 \, \delta_{d1}} F_{2,4}(0).
\label{eq:IRscalingI2}
\end{align}
Eqs.~\eq{eq_alphas} and \eq{eq_alphasdD} then follow by inserting  {\Eqs{IRscalingI1}, \eq{IRscalingI2}}, together with \Eq{small_p}, into \Eq{flow2} and requiring that the exponent of the term leading in the IR on the right matches that on the left-hand side. 

{The coefficients $a_{i}$ 
can be determined 
%=======================================================================
%=======================================================================
\label{app:ai}
by inserting the solutions of Eqs.~\eq{eq_alphas} and \eq{eq_alphasdD} back into Eqs.~\eq{IRscalingI1} and \eq{IRscalingI2} and identifying the leading monomials. }
Equations for the $a_{i}$ are extracted by matching the corresponding pre-factors.
As Eqs.~\eq{eq_alphas} and \eq{eq_alphasdD} have multiple solutions, each case must be handled separately.

We start with $d=1$.
In the following, we use the notation $\delta Z_i'(1) = \left. \text{d} \delta Z_i/\text{d}\hat{p} \right|_{\hat{p}=1}$. 
{Such terms arise in Eqs.~\eq{F11} and \eq{F21} because the respective integrals vanish at $\hat{p}=0$, and the integrand must be Taylor expanded up to leading order in $p$.} The relevant term is proportional to the $\hat p$-derivative of $\delta Z_{i}$.
The solutions of \Eq{eq_alphas} are given in \Eq{alphas}.
If $(\alpha_1,\alpha_2) = (1,5)$ the terms proportional to $F_{1,1}(0)$ and $F_{2,1}(0)$ are dominating. 
The corresponding equations are
\begin{align}
 & a_1(\eta_1-1) = h F_{1,1}(0), \nonumber \\
 & a_2(\eta_2-5) = h (a_1 a_2)^{1/2} F_{2,1}(0),
\end{align}
with
\begin{align}
& F_{1,1}(0) = (8\pi S_1^3 S_2)^{-1} \nonumber \\
& \times \left[2(\eta_1-1)S_1^2 S_2^2 - S_2^2 \delta Z_1'(1) + S_1^2\delta Z_2'(1)\right], \nonumber \\[0.2cm]
& F_{2,1}(0) = (32\pi S_1^6)^{-1} \nonumber \\
& \times\left[4(\eta_1-1)S_1^2 S_2^2 +S_2^2\delta Z_1'(1)+S_1^2\delta Z_2'(1)\right].
\end{align}

If $-2<\alpha_1 < 1$ and $\alpha_2 = -2$, the dominating terms are the ones that are proportional to $F_{1,3}(0)$ and $F_{2,4}(0)$, and thus
\begin{align}
  a_1(\eta_1 - \alpha_1) &= h a_1 a_2 F_{1,3}(0), \nonumber \\
 a_2 (\eta_2+2) &= h a_2^2 F_{2,4}(0),
\label{eq:aclosed_1d}
\end{align}
with
\begin{align}
 & F_{1,3}(0) = F_{2,4}(0) = (8\pi S_1^3 S_2)^{-1}.
\end{align}
While the above equations for the $a_{i}$ are not closed, $h$ can be eliminated by dividing the equations by each other.
This gives the closed equation
\begin{align}
\frac{a_1}{a_2} = - \frac{a_1}{a_2} \frac{\eta_2 +2}{\alpha_1-\eta_1},
\label{eq:ratioaiEqs1d}
\end{align}
equivalent to \Eq{eta12alpha1}.

If $\alpha_1 =-2$ and $\alpha_2 = -2$, the term proportional to $F_{1,3}(0)$ still dominates $\hat{I}_1^{(2)}(\hat{p})$, but the term proportional to $F_{2,3}(0)$ is of the same order as the one proportional to $F_{2,4}(0)$. 
Taking both into account gives
\begin{align}
& a_1 (\eta_1+2) = h a_1 a_2 F_{1,3}(0), \nonumber \\
& a_2 (\eta_2+2) = h \left[ a_1 a_2 F_{2,3}(0)+ a_2^2 F_{2,4}(0)\right],
\end{align}
with
\begin{align}
& F_{2,3}(0) = -S_2 (32 \pi S_1^5)^{-1}. \nonumber \\
\end{align}
Finally, for $\alpha_1 =1$ and $\alpha_2 = -2$ the dominating terms are proportional to $F_{1,1}(0)+ a_1 a_2 F_{1,3}(0)$ and $F_{2,4}(0)$, i.e.,
\begin{align}
& a_1 (\eta_1-1)=  h \left( F_{1,1}(0) + a_1 a_2 F_{1,3}(0)\right), \nonumber \\
& a_2 (\eta_2+2)= h a_2^2 F_{2,4}(0).
\end{align}
As a result, at the end points $(\alpha_1,\alpha_2) \in (\{-2,1\},-2)$, two of the terms on the right hand side of Eqs.~\eq{IRscalingI1} or \eq{IRscalingI2} have the same importance such that the equations for $a_i$ are not the same as for $-2 < \alpha_1 <1$. 
There is no constraint on the value of $\eta_1$ arising there. 
However, under the assumption that $\eta_1$ is a continuous function of $\alpha_1$,  one obtains $\eta_1 = 2-\alpha_1/2$ also at the end points. 
We remark that the calculation reported in Ref.~\cite{Medina1989a} provides a continuum of fixed points, with  $\eta_1 = 2 - \alpha_1/2$ and $0<\alpha_1 <1$, and an additional fixed point with $\eta_1 = 3/2$, the end-point value.
It is not possible to relate $\eta_1$ to $\alpha_i$ at the other solutions of \Eq{alphas} in a similar way. At these points, the full solution of \Eq{flow1} is necessary to verify the existence of the RG fixed points and extract the values of the $\eta_{i}$.

We finally consider the case that {$d \neq 1$}.
\Eq{eq_alphasdD} has the  solutions \eq{alphasdD}.
If $(\alpha_1,\alpha_2) = (1,5)$, we obtain
\begin{align}
 a_1 (\eta_1-1) &= h F_{1,1}(0),\nonumber \\ 
 a_2 (\eta_2-5) &= h (a_1 a_2)^{1/2} F_{2,1}(0), \nonumber \\
\end{align}
with
\begin{align}
&F_{1,1}(0) = 2^d\Omega_d \Gamma(d/2)^2(16\pi (d-1)! \, S_1^3 S_2)^{-1} \nonumber \\
& \quad \times \left[2(\eta_1-1) S_1^2 S_2^2 - S_2^2 \delta Z_1'(1)+S_1^2\delta Z_2'(1)\right], \nonumber \\[0.2cm]
&F_{2,1}(0) = \Omega_d (32d S_1^6)^{-1} \nonumber \\
& \quad \times  \left[4(\eta_1-1)S_1^2 S_2^2+S_2^2\delta Z_1'(1)+S_1^2\delta Z_2'(1)\right].
\end{align}
Here we have introduced the surface factor $\Omega_d = \int_{\Omega} = d \pi^{d/2}[(2\pi)^d \Gamma(d/2+1)]^{-1}$. 
If $0<\alpha_1<1$ and $\alpha_2 = 0$, we get
\begin{align}
 a_1 (\eta_1-\alpha_1) &= h a_1 a_2  F_{1,3}(0), \nonumber \\
 a_2 \, \eta_2 &= h a_2^2 F_{2,4}(0),
\label{eq:aclosed_Dd}
\end{align}
with
\begin{align}
 & F_{1,3}(0) = F_{2,4}(0) = \Omega_d (d-1) (4dS_1^3 S_2)^{-1}.
\end{align}
Again, while the above equations for the $a_{i}$ are not closed, $h$ can be eliminated by dividing the equations by each other.
This gives the closed equation
\begin{align}
\frac{a_1}{a_2} = \frac{a_1}{a_2} \frac{\eta_1-\alpha_1}{\eta_2},
\label{eq:ratioaiEqsDd}
\end{align}
equivalent to \Eq{eta12alpha1}.

If $\alpha_1 = \alpha_2 = 0$, the equations are
\begin{align}
& a_1 \, \eta_1 = h a_1 a_2 F_{1,3}(0), \nonumber \\
& a_2 \, \eta_2 = h \left[ a_1 a_2 F_{2,3}(0)+a_2^2 F_{2,4}(0)\right],
\end{align}
with
\begin{align}
& F_{2,3}(0) = -\Omega_d(d-1)S_2(16dS_1^5)^{-1}.
\end{align}
If $\alpha_1 = 1$ and $\alpha_2 = 0$, one obtains
\begin{align}
 a_1 (\eta_1-1) &= h \left[ F_{1,1}(0) + a_1 a_2 F_{1,3}(0)\right], 
 \nonumber \\
 a_2 \, \eta_2 &= h a_2^2 F_{2,4}(0).
\end{align}
While none of the above sets of equations is closed, we can eliminate $h$ for $(\alpha_1,\alpha_2) \in (]0,1[,0)$, in the same way as before and recover the relation \eq{eta12alpha1}, $2\eta_1 = 2d + 4 -\alpha_1$.

%=======================================================================
\subsection{Scaling limit ($p\gg k$)}
In this subsection, the asymptotic behaviour of the integrals $\hat{I}^{(2)}_{1,2}(\hat{p})$ for $\hat{p}\gg1$ is derived from the respective dependence of the integrals $F_{i,j}(\hat{p})$.
We discuss this for each $F_{i,j}$ separately, taking into account spherical symmetry. 
For $\hat{p}^2 \gg 1$, then also $\hat{q} \sim \hat{p}$, i.e., $\hat{q}^2 - 1 > 0$, and we can approximately set the theta functions and, since $\delta Z_{i}(\hat{q}\to \infty) = 0$, also the $S_{i}(\hat q)$ to one. 
Separating out the leading UV scaling, $F_{i,j}(\hat p\to\infty) \sim \hat p^{\gamma_{i,j}}$, we write the $F_{i,j}$ in the form
\begin{align}
 F_{i,j}(\hat{p}) = \hat{p}^{\gamma_{i,j}} \int_{\Omega} f_{i,j}(1/\hat{p},\hat{\mathbf{p}} \cdot \mathbf{e}_{\mathbf{r}}/\hat{p}).
\label{eq:deff}
\end{align}
The $f_{i,j}$ are finite and non-vanishing at $1/\hat{p} = 0$. 
Note that, in Eqs.~\eq{F12}, \eq{F21} and \eq{F22}, different terms can be leading in the UV such that the above definition of the $f_{i,j}$ {and $\gamma_{i,j}$} depends on the values of the $\eta_{1,2}$. 
Moreover, the denominator of $\T$ in Eqs.~\eq{F12}--\eq{F24} contains a divergence if $\eta_1-\eta_2 > 0$ in which case an additional factor $\hat{p}^{(\eta_2-\eta_1)/2}$ appears. 
This can be seen by recalling the definition $T(\hat q) = (z_1/z_2)^{1/2} k^{2}(\nukk+\nukq)^{-1} $, which gives (recall $\hat q=|\hat{\mathbf{p}}-\mathbf{e}_\mathbf{r}|$) the large-$\hat{p}$ asymptotic behaviour
\begin{align}
T(|\hat{\mathbf{p}}-\mathbf{e}_\mathbf{r}|) 
& \cong \left[\left(\hat{p}^{2} + 1 
-{2}\hat{p} \frac{\mathbf{p}}{p}\cdot\mathbf{e}_{\mathbf{r}}\right)^{(\eta_1-\eta_2)/4}
+\frac{S_1(1)}{S_2(1)}\right]^{-1} 
\nonumber \\
& \cong \left\{ \begin{array}{ll} 
\hat{p}^{-(\eta_1-\eta_2)/2} & \text{if } \eta_1-\eta_2 > 0 \\ S_2(1)/S_1(1) & \text{if } \eta_1-\eta_2 < 0 \end{array} \right. .
\label{eq:Tq}
\end{align}
Having identified the leading scaling behaviour, the integrals can be computed in the limit $\hat p\to \infty$ by neglecting sub-leading contributions to the integrands. 
We can approximate $f_{i,j}(1/\hat{p},\hat{\mathbf{p}} \cdot \mathbf{e}_{\mathbf{r}}/\hat{p}) \cong f_{i,j}(0,\hat{\mathbf{p}} \cdot \mathbf{e}_{\mathbf{r}}/\hat{p})$ in the integrands and perform the angular integration which gives,
for those integrals where $f_{i,j}(0,y)$ does not depend on $y=\hat{\mathbf{p}} \cdot \mathbf{e}_{\mathbf{r}}/\hat{p}$, a surface factor $\Omega_d = \int_{\Omega} = d \pi^{d/2}[(2\pi)^d \Gamma(d/2+1)]^{-1}$.
The asymptotic behaviour of the integrals $F_{1,1}(\hat{p})$, $F_{1,3}(\hat{p})$, and $F_{2,3}(\hat{p})$ can be derived in this way. 
The result is ($S_i \equiv S_i(1)$)
\begin{align}
F_{1,1}(\hat{p}\to&\,\infty) 
\cong \Omega_d S_2^2 \left[\delta_{d1}/2+(d-1)/d\right] \hat{p}^{1-2 \delta_{d1}}
\nonumber\\
&\quad\,\times \hat p^{-(\eta_1+\eta_2)/2},
\\
F_{1,3}(\hat{p}\to&\,\infty) 
 \cong  \frac{\Omega_d}{2 S_1 S_2} \hat{p}^{2-2\delta_{d1}} \nonumber \\
& \times\left\{ \begin{array}{ll} \hat{p}^{-\eta_1} & \text{if } \eta_1>\eta_2 \\[0.1cm] 
\hat{p}^{-\eta_1}(1+S_1/S_2)^{-1} & \text{if } \eta_1 = \eta_2 \\[0.1cm]
 \hat{p}^{-(\eta_1+\eta_2)/2} (S_2/S_1) & \text{if } \eta_1<\eta_2 \end{array} \right.,
\\
F_{2,3}(\hat{p}\to&\,\infty) 
\cong  -\frac{\Omega_d}{2S_1S_2} \hat{p}^{2-2\delta_{d1}} \nonumber \\
& \times \left\{ \begin{array}{ll} \hat{p}^{-2\eta_1+\eta_2} & \text{if } \eta_1>\eta_2 \\[0.1cm]
 \hat{p}^{-\eta_1} \left(1+S_1/S_2\right)^{-3} & \text{if } \eta_1 = \eta_2 \\[0.1cm] 
\hat{p}^{-(\eta_1+\eta_2)/2} \left({S_2/S_1}\right)^{3} & \text{if } \eta_1<\eta_2 \end{array} \right. .
\end{align}

The calculation of the asymptotic behaviour of the integrals \eq{F12}, \eq{F21} and \eq{F22} can become more involved. Two possibilities arise. 
If $\eta_1+\eta_2 \geq -2$, the asymptotic behaviour is determined in the same way as for $F_{1,1}$, $F_{1,3}$ and $F_{2,3}$. 
However, for $\eta_1+\eta_2 < -2$ the leading term of $f_{i,j}(\epsilon\to0,\hat{\mathbf{p}} \cdot \mathbf{e}_{\mathbf{r}}/\hat{p})$ is proportional to $y =\hat{\mathbf{p}} \cdot \mathbf{e}_{\mathbf{r}}/\hat{p}$, and thus vanishes under the angular integral.
In this case, the asymptotically leading term is obtained by expanding  $yT(\hat q)\equiv yT(\hat p,y)$ to order $y^{2}$ before the limit $\hat{p}\to \infty$ is taken and the term that is linear in $y$ is neglected. 
This ensures that we only consider terms that contribute to the angular integration. 
One can check that truncating at order $y^2$ does not affect the asymptotic behaviour. Indeed $y$ enters through the combination $\hat{p}^2-2\hat{\mathbf{p}} \cdot \mathbf{e}_{\mathbf{r}} = (1-2y/\hat{p})\hat p^2$. 
We see that the term of order $y^n$ is multiplied by $1/\hat p^n$ and can only dominate in the asymptotic regime if all the lower order terms are irrelevant.

We discuss the procedure for $F_{1,2}(\hat{p})$ and state the results for the two remaining integrals {$F_{2,1}(\hat{p})$ and $F_{2,2}(\hat{p})$.}
To simplify the derivation we use that $(\eta_{1}+\eta_{2})/2=2\eta_{1}-2-d$ from \Eq{consistency}.
We start by approximating $\delta Z_{i}(\hat{q})\simeq0$, $\theta \left( \hat{q}^2-1\right)=1$ in \Eq{F12}, which gives, defining $\epsilon=1/\hat{p}$ such that $\hat{\mathbf{p}} \cdot \mathbf{e}_{\mathbf{r}} = y/\epsilon$,
\begin{align}
& F_{1,2}(\hat{p}) \cong \int_{\Omega} \left[(S_1 S_2)^{-1}  (\epsilon y-1) - \hat{q}^{-2\eta_1+2+d} \epsilon  y\right] \T ,
\end{align}
with $\hat{q} = \sqrt{1+\epsilon^2-2\epsilon y}/\epsilon$. 
We factor out $\epsilon^{-2\eta_1+2+d}$ from $\hat{q}^{-2\eta_1+2+d}$ in the second term:
\begin{align}
& F_{1,2}(\hat{p}) \cong \int_{\Omega} \, \T \left[ (S_1S_2)^{-1} \left(\epsilon y -1 \right) 
\vphantom{y(1+\epsilon^2-2\epsilon y)^{(-2\eta_1+2+d)/2}}\right. 
\nonumber \\
& \left. \quad -\ \epsilon^{-d-1+2\eta_1}  y(1+\epsilon^2-2\epsilon y)^{(-2\eta_1+2+d)/2} \right] .
\label{eq:befored+2}
\end{align}
The asymptotic behaviour of $T(\hat q)$ is determined by the sign of $-(\eta_{1}-\eta_{2})/2=\eta_{1}-2-d$, see \Eq{Tq}.
For both signs, different $\eta_{1}$ will render either of the terms in \Eq{befored+2} dominating for large $\hat p$ ($\epsilon\to0$).

1. $\eta_1<d+2$, $T(\hat{q}\to\infty) \sim \hat{p}^{\eta_1-d-2}$: 
We write $\T = \epsilon^{-\eta_1+2+d} \, \twidlT$ such that $\tilde{T}(\epsilon\to 0) = 1$ and
\begin{align}
& F_{1,2}(\hat{p}) \cong \int_{\Omega} \left[\epsilon^{-\eta_1+d+2} (S_1S_2)^{-1} \left(\epsilon y-1\right) \right. \nonumber \\
& \quad \left. -\ \epsilon^{\eta_1+1} (1+\epsilon^2-2\epsilon y)^{(-2\eta_1+2+d)/2} y\right] \twidlT .
\label{eq:before(d+1)_2}
\end{align}
There are three sub-cases to be distinguished: 
(a) For $2\eta_1<d+1$, the second term, providing an extra scaling factor $\epsilon^{\eta_1+1}$, is dominant.
Then the leading exponent defined in \eq{deff} reads $\gamma_{1,2}=-\eta_1-1$, and the integrand 
\begin{align}
& f_{1,2}(\epsilon,y) = \left[\epsilon^{-2\eta_1+d+1}(S_1S_2)^{-1} \left(\epsilon y-1\right) \right. \nonumber \\
& \quad \left. -(1+\epsilon^2-2\epsilon y)^{(-2\eta_1+2+d)/2} y \right] \twidlT .
\end{align}
The leading term $f_{1,2}(0,y)= -y$ does not contribute to the angular integral.
Taking the sub-leading factors into account by expanding  to second order in $y$,
\begin{align}
 f_{1,2}&(\epsilon,y) \cong  -\twidlT \Big( y [1+\epsilon^2]^{(-2\eta_1+2+d)/2}  \nonumber\\
& + y^2 \epsilon [1+\epsilon^2]^{(-2\eta_1+d)/2} \Big[ 2\eta_1 -2-d \nonumber \\
& \quad \left.- (\eta_1-2-d)(1+\epsilon^2)^{(-\eta_1+2+d)/2} \, \twidlT \right] \nonumber \\
& + \epsilon^{-2\eta_1+d+1}/(S_1S_2) \Big\{ 1-\epsilon y \nonumber \\
& \qquad \times \left[1 + (\eta_1-2-d) (1+\epsilon^2)^{(-\eta_1+d)/2} \, \twidlT \right] \nonumber \\
& \quad + \epsilon^2 y^2 (\eta_1-2-d) (1+\epsilon^2)^{(-\eta_1-2+d)/2} \, \twidlT \nonumber \\
& \qquad \times \left[ (\eta_1-2-d)(1+\epsilon)^{(-\eta_1+2+d)/2} \, \twidlT \right. \nonumber \\
& \qquad \quad \left. +(d-\eta_1)/2 + 1+ \epsilon^2 \vphantom{(-\eta_1+2+d)(1+\epsilon)^{(-\eta_1+2+d)/2} \, \twidlT}\right] \Big\} \Big),
\end{align}
we find that 
two terms are competing, requiring a further case distinction:
If $\eta_1 < d/2$, the contributions $\propto\epsilon^{-2\eta_1+d+1}$ are sub-leading and the quadratic term in $y$ dominates. 
If $\eta_1 > d/2$, the term not depending on $y$ dominates. 
Both must be account for if $\eta_1 = d/2$.
Then
\begin{align}
& {f}_{1,2}(\epsilon\to0,y) 
 \simeq -\ y -\frac{1}{S_1 S_2} \epsilon \nonumber \\
& \times \left\{ \begin{array}{ll} y^2  \, \eta_1 S_1 S_2 &  \eta_1 < d/2 \\[0.1cm]
  \left(1+ y^2 d S_1 S_2/2 \right) &  \eta_1 = d/2 \\[0.1cm]
 \epsilon^{-2\eta_1+d} &  d/2 < \eta_1 %\\[0.1cm]
< (d+1)/2 \end{array} \right.,
\end{align}
and, after angular integration, $\int_{\Omega} y^{2} = \Omega_d/d$,
\begin{align}
& F_{1,2}(\hat{p}\to\infty) \simeq  -\frac{\Omega_d}{S_1 S_2} \hat{p}^{-2} \nonumber \\[0.1cm]
& \times \left\{ \begin{array}{ll} \hat{p}^{-\eta_1} \, \eta_1 S_1 S_2/d & \eta_1 < d/2 \\[0.1cm] 
 \hat{p}^{-d/2} \left(1+S_1 S_2 /2 \right) & \eta_1 = d/2 \\[0.1cm] 
\hat{p}^{\eta_1-d} & d/2 < \eta_1 %\\[0.1cm]
< (d+1)/2 \end{array} \right. .
\label{eq:asymptotic1}
\end{align}
(b)  For $2\eta_1 = d+1$, both terms under the integral \eq{before(d+1)_2} are equally important. 
We obtain $\gamma_{1,2} = -(d+3)/2$ and
\begin{align}
& f_{1,2}(\epsilon,y) = \left[\frac{\epsilon y-1}{S_1 \, S_2} 
- (1+\epsilon^2-2\epsilon y)^{1/2} y \right]\twidlT.
\end{align}
The relevant contribution is $f_{1,2}(0,y) = - y - (S_1S_2)^{-1}$ while  the terms of order $y^2$ are sub-dominant. 
As a result, the asymptotics \eq{asymptotic1} is supplemented with
\begin{align}
& F_{1,2}(\hat{p}\to\infty) \simeq  - \frac{\Omega_d}{S_1S_2} \hat{p}^{-(d+3)/2} &  \mbox{if}\ \eta_1 = (d+1)/2 .
\label{eq:F12caseb}
\end{align}
(c) For $(d+1)/2<\eta_1<d+2$, the leading terms are interchanged. 
\Eq{before(d+1)_2} provides $\gamma_{1,2} = \eta_1-d-2$ and
\begin{align}
& f_{1,2}(\hat{p}) = \big[\left(\epsilon y-1\right)(S_1 S_2)^{-1}  \nonumber \\
&  \quad -\ \epsilon^{2\eta_1-d-1}(\epsilon^2+1-2\epsilon y)^{(d+2-2\eta_1)/2} y\big] \twidlT.
\end{align}
We find $f_{1,2}(0,y) = -(S_1S_2)^{-1}$, and, together with relation \eq{F12caseb}, the last case of the asymptotics \eq{asymptotic1} reads
\begin{align}
& F_{1,2}(\hat{p}\to\infty) \simeq  -\frac{\Omega_d}{S_1 S_2} \hat{p}^{\eta_1-d-2},\ \mbox{for}\ d/2 < \eta_1 < d+2 .
\label{eq:asymptotic3}
\end{align}

2. $\eta_1 \geq d+2$,  $T(\hat{q}\to \infty) \sim \text{const.}$: 
Now, $\hat{q}^{-\eta_1+2+d}$ is finite for $\epsilon \to 0$, such that no powers of $1/\hat p$ arise from $\T$. 
Two competing terms in  \eq{befored+2} imply three sub-cases.
However, for $\eta_1 \geq d+2$, the term  $\propto\epsilon^{-d-1+2\eta_1}$ is sub-dominant and can be neglected.
The term $\propto(S_{1}S_{2})^{-1}$ in \eq{befored+2}
is dominant, such that $\gamma_{1,2} = 0$ and
\vspace{-0.2cm}
\begin{align}
& f_{1,2}(\epsilon,y) = \big[(\epsilon y-1)(S_1 S_2)^{-1}  \nonumber \\
&  \quad -\epsilon^{2\eta_1-d-1}(1+\epsilon^2-2\epsilon y)^{(d+2-2\eta_1)/2} y\big] \T.
\end{align}
Taking the limit $f_{1,2}(\epsilon\to0,y)$ and performing the angular integral one obtains the final asymptotics
\vspace{-0.2cm}
\begin{align}
& F_{1,2}(\hat{p}\to\infty) \simeq  -\frac{\Omega_d}{S_1 S_2} \hat{p}^{-2} \nonumber \\[0.1cm]
& \times \left\{ \begin{array}{ll} \hat{p}^{-\eta_1} \, \eta_1 S_1 S_2/d & \eta_1 < d/2 \\[0.1cm] 
 \hat{p}^{-d/2} \left(1+S_1 S_2 /2 \right) & \eta_1 = d/2 \\[0.1cm] 
\hat{p}^{\eta_1-d} &  d/2 < \eta_1 
< d+2 \\[0.1cm] 
 \hat{p}^{2} \, (1+S_1/S_2)^{-1} &  \eta_1 = d+2 \\[0.1cm] 
 \hat{p}^{2} S_2/S_1 &  d+2 < \eta_1 \end{array} \right. .
\label{eq:asymptotic4}
\end{align}\postdisplaypenalty=10000
Using analogous arguments we find
\begin{align}
& F_{2,1}(\hat{p}) \sim \frac{\Omega_d}{S_1 S_2} \hat{p}^{-6} \nonumber \\[0.1cm]
& \times \left\{ \begin{array}{ll} \hat{p}^{\eta_1-2d} S_1 S_2 (\eta_1+2+d)/(2d) & \eta_1 < d/2 \\[0.1cm] 
 \hat{p}^{-3d/2} \left[1+ S_1 S_2 (3d+4)/(4d) \right] & \eta_1 = d/2 \\[0.1cm] 
\hat{p}^{3\eta_1-3d} &  d/2 < \eta_1 
< d+2 \\[0.1cm] 
 \hat{p}^{6} \left(1+S_1/S_2\right)^{-3} &  \eta_1 = d+2 \\[0.1cm] 
 \hat{p}^{6} (S_2/S_1)^3 &  d+2 < \eta_1 \end{array} \right.\!\!,
\\ \nonumber
& F_{2,2}(\hat{p}) \sim \frac{\Omega_d}{S_1 S_2} \hat{p}^{-6} \nonumber \\[0.1cm]
& \times \left\{ \begin{array}{ll} \hat{p}^{\eta_1-2d} \, S_1 S_2 (5\eta_1-8-4d)/d & \eta_1 < d/2 \\[0.1cm] 
 \hat{p}^{-3d/2} \left[1 - S_1 S_2 (16+3d)/(2d) \right] & \eta_1 = d/2 \\[0.1cm] 
 \hat{p}^{3\eta_1-3d} &  d/2 < \eta_1 
 < d+2 \\[0.1cm] 
 \hat{p}^{6} \left(1+S_1/S_2\right)^{-3} &  \eta_1 = d+2 \\[0.1cm] 
 \hat{p}^{6} (S_2/S_1)^3 &  d+2 < \eta_1 
 \end{array} \right. \!.%\!.
 \\ \nonumber
\end{align}
\vfill
\pagebreak
\noindent The UV leading terms in of $F_{i,j}$  inserted back into Eqs.~\eq{rescaled_flow_1} and \eq{rescaled_flow_2} allow to compute the asymptotics of $\hat{I}_{1,2}^{(2)}(\hat{p})$, separately for each case. 
The integrals on the right-hand side of \Eq{flow3} converge if $\beta_i < \eta_i$, corresponding to the range \eq{range} for $\eta_{1}$.
With \Eq{consistency}, we find $\hat{I}_{i}^{(2)}(\hat{p}\gg 1) \sim \hat{p}^{\beta_i}$ with
\begin{align}
 \beta_1 
 & = \left\{ 
 \begin{array}{ll} 
 4-2\delta_{d1}-2\eta_1+d 
 &\mbox{if}\ (6+3d-2\delta_{d1})/5 
 \geq \eta_1 
 \\[0.1cm]
 3\eta_1-2d-2 
 & \mbox{if}\ (6+3d-2\delta_{d1})/5 
 \\[0.1cm]
 & \; \; < \eta_1 \leq d+2 
 \\[0.1cm]
 2\eta_1-d 
 &\mbox{if}\ \eta_1 > d+2 
\end{array} \right. ,
\label{eq:beta1}
\\
 \beta_2 
 & = \left\{ \begin{array}{ll} 
 3(\eta_1 -d-2) 
 &\mbox{if}\  \eta_1 \leq d/2 
 \\[0.1cm]
5\eta_1-4d-6 
&\mbox{if}\  d/2 < \eta_1 \leq d+2 
\\[0.1cm]
3\eta_1-2d -2 
&\mbox{if}\  d+2 < \eta_1 
\\[0.1cm]
& \; \; \leq d+2 + 2 \delta_{d1} 
\\[0.1cm] 
4\eta_1-3d-4 -2\delta_{d1} 
&\mbox{if}\  d+2 + 2 \delta_{d1} < \eta_1 
\end{array} \right. .
\label{eq:beta2}
\end{align}

%\bibliography{Master}

%=======================================================================
%=======================================================================

%

\end{document}